\documentclass[12pt,english]{article}
\usepackage[T1]{fontenc}
\usepackage[latin9]{inputenc}
\usepackage[left=1in,top=1in,right=1in,bottom=1in]{geometry}
\usepackage{babel}
\usepackage{amsmath}
\usepackage{amsthm}
\usepackage[pctexhp]{graphicx}
\usepackage{setspace}
\usepackage[numbers,square]{natbib}
\usepackage[unicode=true]
{hyperref}

\makeatletter

\providecommand{\tabularnewline}{\\}

\numberwithin{equation}{section}
\numberwithin{figure}{section}

\usepackage{etex}
\usepackage{pstricks}
\usepackage{pstricks-add,psfrag,pst-node}

\usepackage{tikz}
\usetikzlibrary{tikzmark}

\usepackage{amsmath}
\usepackage{amsthm,thmtools}
\usepackage{amssymb}
\usepackage{graphicx}
\usepackage{latexsym}
\usepackage{authblk}

\newtheorem{theorem}{Theorem}

\declaretheorem[name=Remark]{remark}
\declaretheorem[name=Assumption]{assume}

\newcommand{\norm}[1]{\left\lVert#1\right\rVert}

\makeatother

\begin{document}

\title{Targeted Learning with Daily EHR Data}

\author[1,2]{Oleg Sofrygin\thanks{Oleg Sofrygin is the corresponding author, sofrygin@berkeley.edu}}
\author[1]{Zheng Zhu}
\author[1]{Julie A Schmittdiel}
\author[1]{Alyce S. Adams}
\author[1]{Richard W. Grant}
\author[2]{Mark J. van der Laan}
\author[1]{Romain Neugebauer}

\affil[1]{\footnotesize   Division of Research,  Kaiser Permanente,
  Northern California, Oakland, CA, U.S.A.}
\affil[2]{\footnotesize   Division of Biostatistics, School of Public Health, UC
  Berkeley, USA}

\maketitle
\begin{abstract}

Electronic health records (EHR) data provide a cost and time-effective
opportunity to conduct cohort studies of the effects of multiple time-point
interventions in the diverse patient population found in real-world clinical
settings. Because the computational cost of analyzing EHR data at daily
(or more granular) scale can be quite high, a pragmatic approach has been
to partition the follow-up into coarser intervals of pre-specified length.
Current guidelines suggest employing a 'small' interval, but the feasibility
and practical impact of this recommendation has not been evaluated and
no formal methodology to inform this choice has been developed. We start
filling these gaps by leveraging large-scale EHR data from a diabetes study
to develop and illustrate a fast and scalable targeted learning approach
that allows to follow the current recommendation and study its practical
impact on inference. More specifically, we map daily EHR data into four
analytic datasets using 90, 30, 15 and 5-day intervals. We apply a semi-parametric
and doubly robust estimation approach, the longitudinal TMLE, to estimate
the causal effects of four dynamic treatment rules with each dataset, and
compare the resulting inferences. To overcome the computational challenges
presented by the size of these data, we propose a novel TMLE implementation,
the 'long-format TMLE', and rely on the latest advances in scalable data-adaptive
machine-learning software, xgboost and h2o, for estimation of the TMLE
nuisance parameters.
\end{abstract}
keywords: Big data; Causal inference; Dynamic treatment regimes; EHR; Machine
learning; Targeted minimum loss-based estimation.

\clearpage
\section{Introduction}

The availability of linked databases and compilations of electronic health
records (EHR) has enabled the conduct of observational studies using large
representative population cohorts. This data typically provides information
on the nature of clinical visits (e.g, ambulatory, emergency department,
email, telephone, acute inpatient hospital stay), medication dispensed,
diagnoses, procedures, laboratory test results and any other information
that is continuously generated from patients' encounters with their healthcare
providers. For instance, EHR-based cohort studies have been used to estimate
the relative effectiveness of time-varying interventions in real-life clinical
settings.

The advances in causal inference have provided a sound methodological basis
for designing observational studies and assessing the validity of their
findings. For example, the ``new user design'' \citep{ray2003} advocates
for applying the same rigor and selection criteria used in RCT design to
EHR-based observational studies \citep{hernan2008,hernan2016}. Moreover,
advances in semi-parametric and empirical process theory have allowed for
flexible data-adaptive estimation methods that can incorporate machine
learning into analyses of comparative effectiveness. By lowering the risk
of model misspecification, these data-adaptive approaches can further strengthen
the validity of evidence based on observational studies. Finally, some
of these semi-parametric approaches also allow drawing valid inference
based on formal asymptotic results. For example, the recently proposed
Targeted Minimum Loss-Based Estimation (TMLE) for longitudinal data \citep{vanderLaan:Gruber12a,petersen2014}
-- a doubly robust and locally-efficient substitution estimator.

While recent methodological advances have significantly improved the potential
strength of evidence from observational studies, the practical tools for
conducting such analyses have not kept up with the growing size of EHR
data. In particular, implementation and application of machine learning
to large scale EHR data has proved to be challenging \citep{gruber2015,neugebauer2013}.
EHR data typically includes almost continuous event dates (e.g., data is
updated daily), rather than the discrete event dates from interval assessments
more common in epidemiologic cohort studies and many RCTs (e.g., data is
updated every 3 months). To mitigate the high computing cost of analyzing
EHR data at the daily (or more granular) scale, an analyst typically discretizes
study follow-up by choosing a small number of cutoff time points. These
cutoffs determine the duration of each follow-up time interval and the
total number of analysis time points. The granular EHR data on each subject
is then aggregated into interval-specific measurements for downstream analysis.

Current literature suggests choosing a small time interval \citep{hernan2016}
to define evenly spaced cutoff time points, however there are no clear
guidelines for deciding on the optimal duration of this interval (referred
to as the 'time unit' from hereon). Moreover, in practice, the effect of
selecting different time unit on causal inferences has not been previously
examined within the same EHR cohort. Notably, the choice of a time unit
is often driven by the computational complexity of the estimation procedure,
as much as the subject-specific domain knowledge \citep{neugebauer2013}.
For example, in \citet{neugebauer2014} the authors applied longitudinal
TMLE for estimating the comparative effectiveness of four dynamic treatment
regimes by coarsening the daily EHR data into the 90-day time unit. However,
such coarsening introduces measurement error, which can in turn lead to
bias in the resulting effect estimates. For example, the treatment level
assigned to a patient for one 90-day time-interval might misrepresent the
actual treatment experienced. Intuitively, analyzing data as it is observed
(using the original event dates) should improve causal inferences by avoiding
the reliance on arbitrary coarsening algorithms.

In this paper, we propose a fast and scalable targeted learning implementation
for estimating the effects of complex treatment regimes using EHR data
coarsened with a time unit that can more closely (compared to current practice)
approximate the original EHR event dates. We demonstrate the feasibility
of the proposed approach and evaluate how the choice of progressively larger
time units may effect inference by re-analyzing EHR data from a large diabetes
comparative effectiveness study described in \citet{neugebauer2014}. We
used the granular EHR data generated from the patient's encounters with
the healthcare system and discretized the patient-specific daily follow-up
by mapping it into equally-sized time bins. In separate analyses, the time
unit was varied from 90 days, down to 30, 15 and 5 days. These four time-units
yielded four analytic datasets, each based on the same pool of subjects,
but with a different level of follow-up coarsening as defined by selected
time-unit. Notably, the 5-day time-unit produced a dataset that was nearly
a replica of the original granular EHR dataset. We then applied an analogue
of the double robust estimating equation method first proposed by \citet{Bang:Robins05},
similar to the TMLE described in \citet{vanderLaan:Gruber12a}, to each
of these four datasets. We also compared our results to those obtained
from the previous TMLE analysis based on 90-day time-unit in \citet{neugebauer2014}.

The 90, 30, 15 and 5-day time-unit resulted in datasets with roughly 0.62,
1.81, 3.59 and 8.23 million person-time observations for the entire duration
of the follow-up, respectively. The large number of person-time observations
and the high computational complexity of our chosen estimation procedures
required developing novel statistical software. We carried out our analysis
by implementing a new R package, $\mathtt{stremr}$ \citep{R-stremr},
which streamlines the analysis of comparative effectiveness of static,
dynamic and stochastic interventions in large-scale longitudinal data.
As part of the $\mathtt{stremr}$ R package, we have implemented a computationally
efficient version of the longitudinal TMLE, to which we refer as the long-format
TMLE. Furthermore, for estimation of the nuisance parameters, we relied
on the latest machine learning tools available in R language \citep{r},
such as the Extreme Gradient Boosting with $\mathtt{xgboost}$ \citep{chen2016xgboost}
and fast and scalable machine learning with $\mathtt{h2o}$ \citep{h2o}.
Both of these packages implement a number of distributed and highly data-adaptive
algorithms designed to work well in large data.

The contributions of this article can be summarized as follows. First,
to the best of our knowledge, this is the first time the performance of
longitudinal TMLE has been evaluated on the same EHR data under varying
discretizations of the follow-up time. Furthermore, we present a novel
and computationally efficient version of the longitudinal TMLE. We also
present a possible application of the new $\mathtt{stremr}$ software which
allowed us to analyze such large scale EHR data. Finally, we hope that
our new software will help advance future reproducible research with EHR
data and will contribute to research on time-unit selection and its effects
on inference.

The remainder of this article is organized as follows. In Section~\ref{sec:question},
we describe our motivating research question. In Section~\ref{sec:Data-and-Modeling}
we formally describe the observed data, our statistical parameter and introduce
a novel implementation of the longitudinal TMLE with data-adaptive estimation
of its nuisance parameters. In Section~\ref{sec:results}, we describe
our analyses, present the benchmarks for computing times with the $\mathtt{stremr}$
R package and present our analyses results. Finally, we conclude with a
discussion in Section~\ref{sec:discussion}. Additional materials and
results are provided in our Web Supplement.

\section{Motivating study: comparative effectiveness of dynamic regimes in diabetes
care \label{sec:question}}

The diabetes study and context that motivated this work was previously
described in \citet{neugebauer2014}. Briefly, it has long been hypothesized
that aggressive glycemic control is an effective strategy to reduce the
occurrence of common and devastating microvascular and macrovascular complications
of type 2 diabetes (T2DM). A major goal of clinical care of T2DM is minimization
of such complications through a variety of pharmacological treatments and
interventions to achieve recommended levels of glucose control. The progressive
nature of T2DM results in frequent revisiting of treatment decisions for
many patients as glycemic control deteriorates. Widely accepted stepwise
guidelines start treatment with metformin, then add a secretagogue if control
is not reached or deteriorates. Insulin or (less frequently) a third oral
agent is the next step. Thus, it is common for T2DM patients to be on multiple
glucose-lowering medications.

Current recommendations specify target hemoglobin A1c of $<7\%$ for most
patients \citep{nathan2006,skyler2008}. However, evidence supporting the
effectiveness of a blanket recommendation is inconsistent across several
outcomes \citep{ray2009,duckworth2009,advance2008,holman2008}, especially
when intensive anti-diabetic therapy is required. The effects of intensive
treatment remain uncertain, and the optimal target levels of A1c for balancing
benefits and risks of therapy are not clearly defined. Furthermore, no
additional major trials addressing these questions are underway.

For these reasons, using the electronic health records (EHR) from patients
of seven sites of the HMO Research Network \citep{vogt2004}, a large retrospective
cohort study of adults with T2DM was conducted to evaluate the impact of
various glucose-lowering strategies on several clinical outcomes. More
specifically, the original analyses were based on TMLE and Inverse Probability
Weighting estimation approaches using EHR data coarsened with the 90-day
time unit to contrast cumulative risks under the following four treatment
intensification (TI) strategies denoted by $d_{\theta}$: 'patient initiates
TI at the first time her A1c level reaches or drifts above $\theta$\%
and patient remains on the intensified therapy thereafter' with $\theta=$7,
7.5, 8, or 8.5. Here, we report on secondary analyses to evaluate the impact
of the same glucose-lowering strategies on the development or progression
of albuminuria, a microvascular complication in T2DM using a novel TMLE
implementation and smaller time units.

\section{Data and Modeling Approaches\label{sec:Data-and-Modeling}}

Below, we first describe the structure of the analytic dataset that results
from coarsening EHR data based on a particular choice of time unit.

\subsection{Data structure and causal parameter\label{sub:data-parameter}}

The observed data on each patient in the cohort consist of measurements
on exposure, outcome, and confounding variables updated at regular time
intervals between study entry and until each patient's end of follow-up.
The time (expressed in units of 90, 30, 15 or 5 days) when the patient's
follow-up ends is denoted by $\tilde{T}$ and is defined as the earliest
of the time to failure, i.e., albuminuria development or progression, denoted
by $T$ or the time to a right-censoring event denoted by $C$. The following
three types of right-censoring events experienced by patients in the study
were distinguished: the end of follow-up by administrative end of study,
disenrollment from the health plan and death. For patients with normoalbuminuria
at study entry, i.e., microalbumin-to-creatinine ratio (ACR) $<$30, we
defined failure as an ACR measurement indicating either microalbuminuria
(ACR 30 to 300) or macroalbuminuria (ACR$>$300). For patients with microalbuminuria
at study entry, we defined failure as an ACR measurement indicating macroalbuminuria.
Addition inclusion and exclusion criteria described in \citet{neugebauer2014}
yielded the final sample size $n=51,179$.

At each time point $t=0,\ldots,\tilde{T}$, the patient's exposure to an
intensified diabetes treatment is represented by the binary variable $A^{T}(t)$,
and the indicator of the patient's right-censored status at time $t$ is
denoted by $A^{C}(t)$. The combination $A(t)=(A^{T}(t),A^{C}(t))$ is
referred to as the action at time $t$. At each time point $t=0,\ldots,\tilde{T}$,
covariates, such as A1c measurements (others are listed in Table I of \citet{neugebauer2015}),
are denoted by the multi-dimensional variable $L(t)$ and defined from
EHR measurements that occur before the action at time $t$, $A(t)$, or
are otherwise assumed not to be affected by the actions at time $t$ or
thereafter, ($A(t),A(t+1),\ldots$). In addition, data collected at each
time $t$ includes an outcome process denoted by $Y(t)$ - an indicator
of failure prior to or at $t$, formally defined as $Y(t)=I(T\leq t)$.
By definition, the outcome is thus missing at $t=\tilde{T}$ if the person
was right-censored at $t$.

To simplify notation, we use over-bars to denote covariate and exposure
histories, e.g., a patient's exposure history through time $t$ is denoted
by $\bar{A}(t)=(A(0),\ldots,A(t))$. We assume the analytic dataset is
composed of $n$ independent and identically distributed (iid) realizations
$(O_{i}:i=1,\ldots,n)$ of the following random variable $O=(\bar{L}(\tilde{T}),\bar{A}(\tilde{T}),\bar{Y}(\tilde{T}-1),(1-A^{C}(\tilde{T}))Y(\tilde{T}))\sim P$.
We also assume that each $O_{i}$ is drawn from distribution $P$ belonging
to some model $\mathcal{M}$. By convention, we extend the observed data
structure using first $Y_{i}(\tilde{T}_{i})=0$ if $A_{i}^{C}(\tilde{T}_{i})=1$
and then $O_{i}(t)=O_{i}(\tilde{T}_{i})$ for $t>\tilde{T}_{i}$. Note
that these added degenerate random variables will not be used in the practical
implementation of our estimation procedure, yet they will allow us to simplify
the presentation in the following section. In particular, this convention
implies that whenever $Y(\tilde{T}_{i})=1$, the outcomes $Y_{i}(t)$ are
deterministically set to $1$ for all $t>\tilde{T}_{i}$.

One common way to store $(O_{1},\ldots,O_{n})$ in a computer is with the
so-called ``long-format'' dataset, where each row contains a record of
a single person-time observation $O_{i}(t)=(L_{i}(t),A_{i}(t),Y_{i}(t))$,
for some $i\in\{1,\ldots,n\}$ and $t\in\{0,\ldots,\tilde{T}_{i}\}$. For
time-to-event data, the long-format can be especially convenient, since
only the relevant (non-degenerate) information is kept, while all degenerate
observation-rows such that $t>\tilde{T}_{i}$ are typically discarded.
An alternative way to store the same analytic dataset is by using the so-called
``wide-format'', which includes values for the degenerate part of the
observed data structure $O_{i}(t)$ for $\tilde{T}_{i}<t\leq K$, where
$K=\max(\tilde{T}_{i}:i=1,...,n)$. For the remainder of this paper we
assume that data are stored in long-format.

In this study, we aim to evaluate the effect of dynamic treatment interventions
on the cumulative risk of failure at a pre-specified time point $t_{0}$.
The dynamic treatment interventions of interest correspond to treatment
decisions made according to given clinical policies for initiation of an
intensified therapy based on the patient's evolving A1c level. These policies
denoted by $d_{\theta}$ were described above. Formally, these policies
are individualized action rules \citep{vanderlaan2007} defined as a vector
function $\bar{d}_{\theta}=(d_{\theta,0},\ldots,d_{\theta,t_{0}})$ where
each function, $d_{\theta,t}$ for $t=0,\ldots,t_{0}$, is a decision rule
for determining the action regimen (i.e., a treatment and right-censoring
intervention) to be experienced by a patient at time $t$, given the action
and covariate history measured up to a given time $t$. More specifically
here, we consider the action rule $(L(t),A(t-1))\mapsto d_{\theta,t}(L(t),A(t-1))\in\{0,1\}\times\{0\}$
as a function for assigning the treatment action $A(t)$. Note that these
rules are restricted to set the censoring indicators $A^{C}(t)=0$. Furthermore,
at each time $t$, we define the dynamic rule $d_{\theta,t}$ by setting
$A^{T}(t)=0$ if and only if the patient was not previously treated with
an intensified therapy (i.e., $A(t-1)=0$) and the A1c level at time $t$
(an element of $L(t)$) was lower than or equal to $\theta$, and, otherwise,
setting $A^{T}(t)=1$. Finally, for each observation $O_{i}$, we define
the treatment process $\bar{A}_{i}^{\theta}(t)=(A_{i}^{\theta}(0),\ldots,A_{i}^{\theta}(t))$
as the treatment sequence that would result from sequentially applying
the previous action rules, i.e., $(d_{\theta,0}(L_{i}(0)),\ldots,d_{\theta,t}(L_{i}(t),A_{i}^{\theta}(t-1)))$,
starting at time $0$ through $t$. Note that the rules defined by $\bar{d}_{\theta}$
are deterministic for a given $O_{i}$ and thus $\bar{A}_{i}^{\theta}$
is fixed conditional on the event $(O_{i}=o_{i})$. For notation convenience,
we also set $\bar{A}_{i}^{\theta}(t)=\bar{A}_{i}^{\theta}(\tilde{T}_{i})$,
for $t>\tilde{T}_{i}$, noting that, in practice, $\bar{A}_{i}^{\theta}(t)$
will be evaluated only when $t\leq\tilde{T}_{i}$ and can be set as arbitrary
for all other $t$.

Suppose $Y_{\bar{d}_{\theta}}(t_{0})$ for $\theta=\theta_{1},\theta_{2}$
denotes a patient's potential outcome at time $t_{0}$ had she been treated
between study entry and time $t_{0}$ according to the decision rule $\bar{d}_{\theta}$.
Note that the corresponding sequence of treatment interventions is not
necessarily equal to $\bar{A}_{i}^{\theta}(t_{0})$. The parameter of interest
denoted by $\psi^{\theta_{1},\theta_{2}}(t_{0})$ is then defined as the
\emph{causal} risk difference between the cumulative risks of two distinct
dynamic treatment strategies $\bar{d}_{\theta_{1}}$ and $\bar{d}_{\theta_{2}}$
at $t_{0}$: $\psi^{\theta_{1},\theta_{2}}(t_{0})=\psi^{\theta_{1}}(t_{0})-\psi^{\theta_{2}}(t_{0}),$
where $\psi^{\theta}(t_{0})=P(Y_{\bar{d}_{\theta}}(t_{0})=1)$, i.e., the
cumulative risk associated with rule $\bar{d}_{\theta}$ at $t_{0}$. The
above definition of the causal parameter of interest relies on the counterfactual
statistical framework, which is omitted here for brevity. We refer the
reader to earlier work in \citet[Appendices B and D]{neugebauer2012} and
\citet{vanderlaan2007} for a detailed description of the relevant concepts.

\subsection{Identifiability and the statistical parameter of interest}

As discussed in the next paragraph, identifiability of the causal parameter
with the observational data relies on at least two assumptions: no unmeasured
confounding and positivity \citep[Appendix C]{neugebauer2012}. If the
counterfactual outcomes are not explicitly defined based on the more general
structural framework through additional explicit assumptions encoded by
a causal diagram \citep{pearl2009causal}, then an additional consistency
assumption is made \citep{vanderweele2009concerning}.

Without loss of generality, suppose that we are interested in estimating
the cumulative risk $\psi^{\theta}(t_{0})$ at $t_{0}=1$ under fixed dynamic
regimen $\bar{d}_{\theta}=(d_{\theta,0},d_{\theta,1})$. Define $L'(k)=(L(k),\bar{O}(k-1))$,
for $k=0,\ldots,t_{0}$, where by convention $(A(-1),L(-1))$ is an empty
set and $Y(-1)=0$. Let $o=(\bar{l}(\tilde{t}),\bar{a}(\tilde{t}),\bar{y}(\tilde{t}))$
denote a particular fixed realization of $O$. We can now define the following
recursive sequence of expectations:
\[
Q_{1}(a,l)=E_{P}(Y(1)|A(1)=a,L'(1)=l),
\]
\begin{eqnarray*}
Q_{0}(a,l) & = & E_{P}(Q_{1}(A^{\theta}(1),L'(1))|A(0)=a,L'(0)=l)
\end{eqnarray*}
\[
Q_{-1}=E_{P}(Q_{0}(A^{\theta}(0),L'(0)),
\]
where we remind that $(A^{\theta}(0),A^{\theta}(1))$ was previously defined
as the sequence of treatments $A^{\theta}(0)=d_{\theta,0}(L(0))$ and $A^{\theta}(1)=d_{\theta,1}(\bar{L}(1),A^{\theta}(0))$.
Note that, by definition $Q_{1}$ is 1 whenever $Y(0)=1$, since in this
case $Y(1)$ is degenerate. \begin{remark} Each $Q_{k}$, for $k=-1,0$,
is defined by taking the previous conditional expectation, $Q_{k+1}$,
evaluating it at $A^{\theta}(k+1)$ and $L^{'}(k+1)$ and then marginalizing
over the intermediate covariates $L^{'}(k+1)$.\end{remark} Under the
identifiability assumptions mentioned above, we show in Web Supplement
A that the statistical parameter $\Psi^{\theta}(t_{0})(P)=Q_{-1}$ is equal
to the causal cumulative risk $\psi^{\theta}(t_{0})$ for dynamic rule
$\bar{d}_{\theta}$. We note that the above representation of $Q_{-1}$
is analogous to the iterative conditional expectation representation used
in \citet{vanderLaan:Gruber12a}, with one notable difference: our parameter
evaluates $Q_{1}$ and $Q_{0}$ with respect to the latest values of the
counterfactual treatment, $A^{\theta}(1)$ and $A^{\theta}(0)$, respectively.
This is in contrast to the iterative conditional expectations in \citet{vanderLaan:Gruber12a},
where conditioning in $Q_{1}$ would be evaluated with respect to the entire
counterfactual history of exposures $\bar{A}^{\theta}(1)$. As we show
in Web Supplement A, these two parameter representations happen to be equivalent.
However, our particular target parameter representation above will allow
us to develop a TMLE that is computationally faster and more scalable to
a much larger number of time-points.

We introduce the following notation, which will be useful for the description
of the TMLE in next section: let $Q=(Q_{1},Q_{0})$ and, for $k=0,1$,
define the treatment mechanism $g_{A(k)}(a(k)|l'(k))=P(A(k)=a(k)|L^{'}(k)=l'(k))$,
i.e., the conditional probability that $A(k)$ is equal to $a(k)$, conditional
on events $L'(k)$ being set to some fixed history $l'(k)$. Finally, let
$g=(g_{A(0)},g_{A(1)})$.

\subsection{Long-format TMLE for time-to-event outcomes\label{sub:lfTMLE}}

Doubly robust approaches allow for consistent estimation of $\Psi^{\theta}(t_{0})(P)$
even when either the outcome model for $Q$ or the exposure model for $g$
is misspecified. Among the class of doubly robust estimators, those that
are based on the substitution principle, such as the longitudinal TMLE
in \citet{vanderLaan:Gruber12a}, might be preferable, since the substitution
principle may offer improvements in finite sample behavior of an estimator.
The TMLE described in \citet{vanderLaan:Gruber12a} is an analogue of the
double robust estimating equation method presented in \citet{Bang:Robins05}.
While there are several possible ways to implement the longitudinal TMLE
procedure (e.g., \citet{stitelman2011,vanderLaan:Gruber12a,R-ltmle}),
our implemented version, referred to as ``long-format TMLE'', has been
adapted to work efficiently with large scale time-to-event EHR datasets.

As a reminder, for each subject $i$, we defined $L'_{i}(k)$ to include
$i$'s entire covariate history up to time-point $k$, in addition to $L_{i}(k)$
itself. However, due to the curse of dimensionality, estimating $Q_{k}$
based on all $L'(k)$, when $k$ is sufficiently large and $L(k)$ is high-dimensional,
will generally result in a poor finite-sample performance. Thus, to control
the dimensionality , we replace $L'(k)$ with a user-defined summary $f_{k}(L'(k))$,
where $f_{k}(\cdot)$ is an arbitrary mapping $(\bar{L}(k),\bar{A}(k-1))\mapsto R^{d}$
such that $d$ is fixed for all $k=0,\ldots,\max(\tilde{T})$. For example,
in our data analyses presented in the following sections we defined the
mapping $f_{k}(L'(k))$ as $(L(0),L(k),A(k-1))$. This approach allows
the practitioner to control the dimensionality of the regression problem
when fitting each $Q_{k}$ model. Furthermore, by forcing all relevant
confounders for time-point $k$ to be defined in a single person-time row
via the mapping $f_{k}(\cdot)$ (i.e., $(f_{k}(\cdot),A_{i}(k),Y_{i}(k))$),
we can also simplify the implementation of the iterative part of TMLE algorithm
that fits the initial model for $Q_{k}$, as we describe next.

Applying the mappings $f_{k}(\cdot)$ to our observed data on $n$ subjects,
$(O_{1},\ldots,O_{n})$, results in a new, reduced, long-format representation
of the data, where each row of the reduced dataset is defined by the following
person-time observation $(f_{k}(L'_{i}(k)),A_{i}(k),Y_{i}(k))$, for some
$i\in\{1,\ldots,n\}$ and $k\in\{0,\ldots,\tilde{T}_{i}\}$. For example,
for subject $i$ with $\tilde{T}_{i}=0$, the entire reduced representation
of $O_{i}$ consists of just a single person-time row $(f_{0}(L_{i}(0)),A_{i}(0),Y_{i}(0))$.
Our proposed TMLE algorithm, outlined below, will work directly with this
reduced long-format representation of $(O_{1},...,O_{n})$. The algorithm
relies on the representation of $\Psi^{\theta}(t_{0})(P)$ from the previous
section, where one makes predictions based on only the last treatment value
for each time-point. This in turn allows us to keep the input data in the
reduced long-format at all times, substantially lowering the memory footprint
of the procedure.

We now describe the long-format TMLE algorithm for estimating parameter
$\Psi^{\theta}(t_{0})(P)$ indexed by the fixed dynamic regimen $\bar{d}_{\theta}=(d_{\theta,0},...,d_{\theta,t_{0}})$,
where for simplicity, we let $t_{0}=1$. A more detailed description of
this TMLE is also provided in the Web Supplement B. Briefly, the algorithm
proceeds recursively by estimating each $Q_{k}$ in $Q$, for $k=1,0$.
Prior to that, we instantiate a new variable $\tilde{Q}_{(k+1)}=Y(k)$,
for $k=0,\ldots,\tilde{T}$ and we make one final modification to our reduced
long-format dataset by adding a new column of subject-specific cumulative
weight estimates, defined for each row $k$ as $\hat{wt}(k)=\prod_{j=0}^{k}\frac{I(\bar{A}(j)=\bar{A}^{\theta}(j))}{\hat{g}_{A(j)}(O)}$,
where $k=0,\ldots,\tilde{T}$ and $\hat{g}_{A(j)}$ is the estimator of
$g_{A(j)}$ at $j$. For iteration $k=1$, one starts by obtaining an initial
estimate $\hat{Q}_{k}$ of $Q_{k}$ by regressing $\tilde{Q}_{(k+1)}$
against $(A(k),f_{k}(L'(k)))$ based on some parametric (e.g., logistic)
model, for all subjects such that $\tilde{T}\geq k$ and $A^{C}(k)=0$
(i.e., this fit is performed among subjects who were at risk for the event
at time $k$). Alternatively, the regression fit can be obtained by using
a subset of subjects such that $A(k)=A^{\theta}(k)$ or based on a data-adaptive
estimation procedure as discussed later. Next, one estimates the intercept
$\varepsilon^{k}$ with an intercept-only logistic regression for the outcome
$\tilde{Q}_{(k+1)}$ using the offset $\mbox{logit}\hat{Q}_{k}(A^{\theta}(k),f_{k}(L'(k)))$,
and the weights $\hat{wt}(k)$, where $\mbox{logit}(x)=\log\left(\frac{x}{1-x}\right)$.
The last regression defines the TMLE update $\hat{Q}_{k}^{*}$ of $\hat{Q}_{k}$
as $\mbox{expit}\left(\mbox{logit}\hat{Q}_{k}(a^{\theta}(k),f_{k}(l'(k)))+\hat{\varepsilon}^{k}\right)$,
for any realization $(a^{\theta}(k),l'(k))$ and $\mbox{expit}(x)=\frac{1}{1+e^{-x}}$.
Finally, if $k>0$, for all subjects such that $\tilde{T}\geq k$, we compute
the TMLE update, defined as $\hat{Q}_{k}^{*}(A^{\theta}(k),f_{k}(L'(k)))$,
and use this update to over-write the previously defined instance of $\tilde{Q}_{k}$.
Note that $\tilde{Q}_{k}$ remains set to $Y(k-1)$ for all subjects with
$\tilde{T}<k$. The illustration of this over-writing scheme for $\tilde{Q}_{k}$
is also presented in Figure \ref{fig:Schematic-TMLE} for iteration $k=1$,
using a toy example for three hypothetical subjects. The same procedure
is now repeated for iteration $k=0$, using the outcome $\tilde{Q}_{(k+1)}$,
resulting in TMLE update $\hat{Q}_{k}^{*}$, for all subjects $i=1,\ldots,n$.
Finally, the TMLE of $\Psi^{\theta}(t_{0})(P)=Q^{-1}$ is defined as $\hat{\Psi}^{\theta}(t_{0})=\frac{1}{n}\sum_{i=1}^{n}\hat{Q}_{0}^{*}(A_{i}^{\theta}(0),f_{0}(L'_{i}(0)))$.
TMLE estimate of the causal RD $\psi^{\theta_{1},\theta_{2}}(t_{0})$ can
be now evaluated as $\hat{\Psi}^{\theta_{1}}(t_{0})(P)-\hat{\Psi}^{\theta_{2}}(t_{0})(P)$,
where the above described procedure is carried out separately to estimate
$\Psi^{\theta_{1}}(t_{0})(P)$ and $\Psi^{\theta_{2}}(t_{0})(P)$, for
rules $\bar{d}_{\theta_{1}}$ and $\bar{d}_{\theta_{2}}$, respectively.

Note that in above description, each initial estimate $\hat{Q}_{k}$ can
be obtained by either stratifying the subjects based on $A(k)=A^{\theta}(k)$
(referred to as ``stratified TMLE'') or by pooling the estimation among
all subjects at risk of event at time $k$ (referred to as ``pooled TMLE'').
Furthermore, in our data analyses described in Section \ref{sec:results},
we use the stratified TMLE procedure. Finally, the inference can be obtained
using the approach described in our Web Supplement C\textbf{,} based on
the asymptotic results from prior papers.

\begin{figure}
\begin{centering}
1) Input data with initialized $\tilde{Q}_{(t+1)}$ values:
\par\end{centering}

\begin{centering}
\begin{tabular}{|c|c|c|c|c|c|c|c|}
\hline
$i$ & $t$ & $\tilde{T}$ & $f_{t}(L'(t))$ & $A^{C}(t)$ & $Y(t)$ & $\tilde{Q}_{(t+1)}$ & $\hat{Q}_{t}^{*}$\tabularnewline
\hline
\hline
$1$ & $0$ & $0$ & . & 0 & $1$ & $1$ & \textbf{NA}\tabularnewline
\hline
$2$ & $0$ & $1$ & . & 0 & $0$ & $0$ & \textbf{NA}\tabularnewline
\hline
$2$ & $1$ & $1$ & . & 0 & $1$ & $1$ & \textbf{NA}\tabularnewline
\hline
3 & 0 & 1 & . & 0 & 0 & 0 & \textbf{NA}\tabularnewline
\hline
3 & 1 & 1 & . & 1 & \textbf{NA} & \textbf{NA} & \textbf{NA}\tabularnewline
\hline
\end{tabular}
\par\end{centering}

\begin{centering}
\vspace{0.8cm}
2) TMLE updates $\hat{Q}_{k}^{*}$ of $Q_{k}$ evaluated at iteration $k=1$:
\par\end{centering}

\begin{centering}
\begin{tabular}{|c|c|c|c|c|c|c|c|}
\hline
$i$ & $t$ & $\tilde{T}$ & $f_{t}(L'(t))$ & $A^{C}(t)$ & $Y(t)$ & $\tilde{Q}_{(t+1)}$ & $\hat{Q}_{t}^{*}$\tabularnewline
\hline
\hline
$1$ & $0$ & $0$ & . & 0 & $1$ & $1$ & \textbf{NA}\tabularnewline
\hline
$2$ & $0$ & $1$ & . & 0 & $0$ & $0$ & \textbf{NA}\tabularnewline
\hline
$2$ & $1$ & $1$ & . & 0 & $1$ & $1$ & $\hat{Q}_{1,i}^{*}$\tabularnewline
\hline
3 & 0 & 1 & . & 0 & 0 & 0 & \textbf{NA}\tabularnewline
\hline
3 & 1 & 1 & . & 1 & \textbf{NA} & \textbf{NA} & $\hat{Q}_{1,i}^{*}$\tabularnewline
\hline
\end{tabular}
\par\end{centering}

\begin{centering}
\vspace{0.8cm}

\par\end{centering}

\begin{centering}
3) Updated $\tilde{Q}_{(t+1)}$ values at the end of iteration $k=1$:
\par\end{centering}

\begin{centering}
\begin{tabular}{|c|c|c|c|c|c|c|c|}
\hline
$i$ & $t$ & $\tilde{T}$ & $f_{t}(L'(t))$ & $A^{C}(t)$ & $Y(t)$ & $\tilde{Q}_{(t+1)}$ & $\hat{Q}_{t}^{*}$\tabularnewline
\hline
\hline
$1$ & $0$ & $0$ & . & 0 & $1$ & $1$ & \textbf{NA}\tabularnewline
\hline
$2$ & $0$ & $1$ & . & 0 & $0$ & $\hat{Q}_{1,i}^{*}$\tikzmark{b} & \textbf{NA}\tabularnewline
\hline
$2$ & $1$ & $1$ & . & 0 & $1$ & $1$ & \tikzmark{a}$\hat{Q}_{1,i}^{*}$\tabularnewline
\hline
3 & 0 & 1 & . & 0 & 0 & $\hat{Q}_{1,i}^{*}$\tikzmark{d} & \textbf{NA}\tabularnewline
\hline
3 & 1 & 1 & . & 1 & \textbf{NA} & \textbf{NA} & \tikzmark{c}$\hat{Q}_{1,i}^{*}$\tabularnewline
\hline
\end{tabular}
\par\end{centering}

\begin{tikzpicture}[overlay, remember picture, yshift=.25\baselineskip, shorten >=.5pt, shorten <=.5pt]
\draw [->] ([yshift=.75pt]{pic cs:a}) -- ({pic cs:b});
\draw [->] ([yshift=.75pt]{pic cs:c}) -- ({pic cs:d});
\end{tikzpicture}

\begin{centering}
\vspace{0.3cm}

\par\end{centering}

\caption{Illustration of the long-format TMLE updating step for the outcome $\tilde{Q}_{(t+1)}$
at iteration $k=1$ for three subjects $i=1,2,3$. Note that the TMLE update
$\hat{Q}_{1,i}^{*}=\hat{Q}_{1}^{*}(A_{i}^{\theta}(1),f_{1}(L_{i}'(1)))$
is also defined for the censored subject $i=3$ at $k=1$. \label{fig:Schematic-TMLE}}
\end{figure}

\subsection{Data-adaptive estimation via cross-validation\label{sub:super-learner}}

The double-robustness property of the TMLE means that its consistency hinges
on the crucial assumption that at least one of the two nuisance parameters
($g$, $Q$) is estimated consistently. Current guidelines suggest that
the nuisance parameters, such as propensity scores, should be estimated
in a flexible and data-adaptive manner \citep{hernan2016,neugebauer2016,gruber2015}.
However, traditionally in observational studies, these nuisance parameters
have been estimated based on logistic regressions, with main terms and
interaction terms often chosen based on the input from subject matter experts
\citep{schnitzer2014,neugebauer2012,bodnar2004,cole2003,hernan2000}. In
contrast, a data-adaptive estimation procedure provides an opportunity
to learn complex patterns in the data which could have been overlooked
when relying on a single parametric model.

For instance, improved finite-sample performance from data-adaptive estimation
of the nuisance parameters has been previously noted with Inverse Probability
Weighting (IPW) estimation, a propensity score-based alternative to TMLE
\citep{neugebauer2016}.  It has been suggested that problems with parametric
modeling approaches can arise even in studies with no violation of the
positivity assumption. For example, the predicted propensity scores from
the misspecified logistic models might be close to 0 or 1, resulting in
unwarranted extreme weights which could be avoided with data-adaptive estimation.
Similarly, these extreme weights may also lead to finite-sample instability
for doubly-robust estimation approaches, such as the long-format TMLE.
These considerations provide further motivation for the use of the data-adaptive
estimation procedures.

Many machine learning (ML) algorithms have been developed and applied for
data-adaptive estimation of nuisance parameters in causal inference problems
\citep{mccaffrey2004propensity,lee2010improving,westreich2010propensity}.
However, the choice of a single ML algorithm over others is unlikely to
be based on real subject-matter knowledge, since: ``\emph{in practice
it is generally impossible to know a priori which {[}ML procedure{]} will
perform best for a given prediction problem and data set}'' \citet{vanderLaan:polley07}.
To hedge against erroneous inference due to arbitrary selection of a single
algorithm, an ensemble learning approach known as discrete Super Learning
(dSL) \citep{vanderLaan:polley07,polley2011super} can be utilized. This
approach selects the optimal ML procedure among a library of candidate
estimators. The optimal estimator is selected by minimizing the estimated
expectation of a user-specified loss function (e.g., the negative log-likelihood
loss) \citep{laan2006cross}. Cross-validation is used to assess an expected
loss associated with each candidate estimator, which protects against overfitting
and ensures that the final selected estimator (called the 'discrete super
learner') performs asymptotically as well (in terms of the expected loss)
as any of the candidate estimators considered \citep{dudoit2005asymptotics}.
Because of the general asymptotic and finite-sample formal dSL results,
in this work we favor the approach of super learning over other existing
ensemble learning approaches.

Prior applications of super learning in R \citep{R-superlearner} have
noted the high computational cost of aggressive super learning, especially
for large datasets (e.g., dSL library contains a large number of computationally
costly ML algorithms) \citep{gruber2015, neugebauer2013}. Because of these
limitations, which are also compounded by the choice of a smaller time
unit in our study, we implemented a new version of the discrete super learner,
the $\mathtt{gridisl}$ R package \citep{R-gridisl}. This R package is
utilized for estimation of the TMLE nuisance parameters by the R package
$\mathtt{stremr}$. Below, we describe how $\mathtt{gridisl}$ builds on
the latest advances in scalable machine learning software, $\mathtt{xgboost}$
\citep{chen2016xgboost} and $\mathtt{h2o}$ \citep{h2o}, which makes
it feasible to conduct more aggressive super learning in EHR-based cohort
studies with small time units.

In our implementation of the discrete super learner, we focus on the negative
log-likelihood loss function. The candidate machine learning algorithms
that can be included in our ensembles are distributed high-performance
$\mathtt{xgboost}$ and $\mathtt{h2o}$ implementations of the following
algorithms: random forests (RFs), gradient boosting machines (GBMs) \citep{friedman2001greedy,chen2016xgboost,h2o_GBM_booklet},
logistic regression (GLM), regularized logistic regression, such as, LASSO,
ridge and elastic net \citep{zou2005regularization,friedman2010regularization}.
GBM is an automated and data-adaptive algorithm that can be used with large
number of covariates to fit a flexible non-parametric model. For overview
of GBMs we refer to \citet{hastie09elements}. The advantage of procedures
like classification trees and GBM is that they allow us to search through
a large space of model parameters, accounting for the effects of many covariates
and their interactions, thereby reducing bias in the resulting estimator
regardless of the distribution of the data that defined the true values
of the nuisance parameters. The large number of possible tuning parameters
available for the estimation procedures in our ensemble required conducting
a grid search over the space of such parameters. We note that $\mathtt{gridisl}$
leverages the internal cross-validation implemented in $\mathtt{h2o}$
and $\mathtt{xgboost}$ R packages for additional computational efficiency.

\section{Analysis\label{sec:results}}

\begin{center}
\begin{table}[th]
\begin{centering}
\begin{tabular}{|c|c|c|c|c|c|}
\hline
time-unit (day) & $n$ for $\hat{g}$ & GLM $\hat{g}$ & dSL $\hat{g}$ & GLM $\hat{Q}^{*}$ & dSL $\hat{Q}^{*}$\tabularnewline
\hline
\hline
90 & 0.6M & 0.05 & 4.95 & 0.06 & 0.81\tabularnewline
\hline
30 & 1.8M & 0.10 & 14.69 & 0.17 & 2.38\tabularnewline
\hline
15 & 3.6M & 0.17 & 28.03 & 0.34 & 4.67\tabularnewline
\hline
5 & 8.2M & 0.41 & 13.43 & 1.11 & 13.94\tabularnewline
\hline
\end{tabular}
\par\end{centering}

\caption{Benchmarks for $\mathtt{stremr}$ with compute time for $\hat{g}$ and
TMLE $\hat{Q}^{*}$ reported separately for parametric approach with logistic
main-term models (GLM $\hat{g}$ and GLM $\hat{Q}^{*}$) and data-adaptive
approach with discrete super learning (dSL $\hat{g}$ and dSL $\hat{Q}$).
The running times are displayed in hours.\label{tab:Benchmarks}}
\end{table}

\par\end{center}

\begin{figure}
\begin{centering}
\includegraphics[scale=0.6]{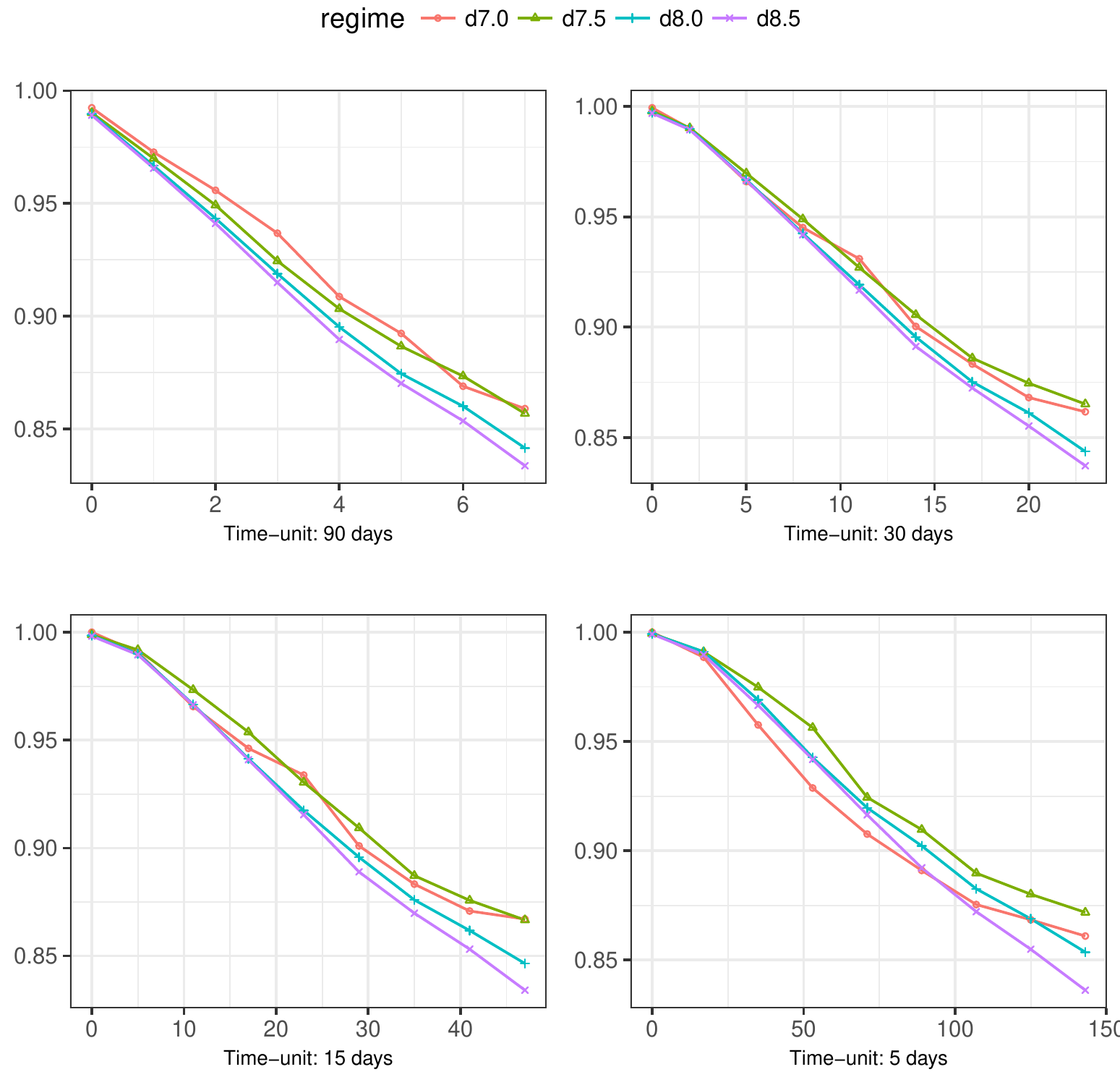}
\par\end{centering}

\centering{}\caption{TMLE survival estimates for data-adaptive modeling approach for $Q$ and
$g$, contrasting four dynamic interventions $(d_{7.0},d_{7.5},d_{8.0},d_{8.5})$
for 90 day (top-left panel), 30 day (top-right panel), 15 day (bottom-left
panel) and 5 day (bottom-right panel) time-unit over two years of follow-up.\label{fig:surv-TMLE-SL}}
\end{figure}

\begin{figure}
\begin{centering}
\includegraphics[scale=0.68]{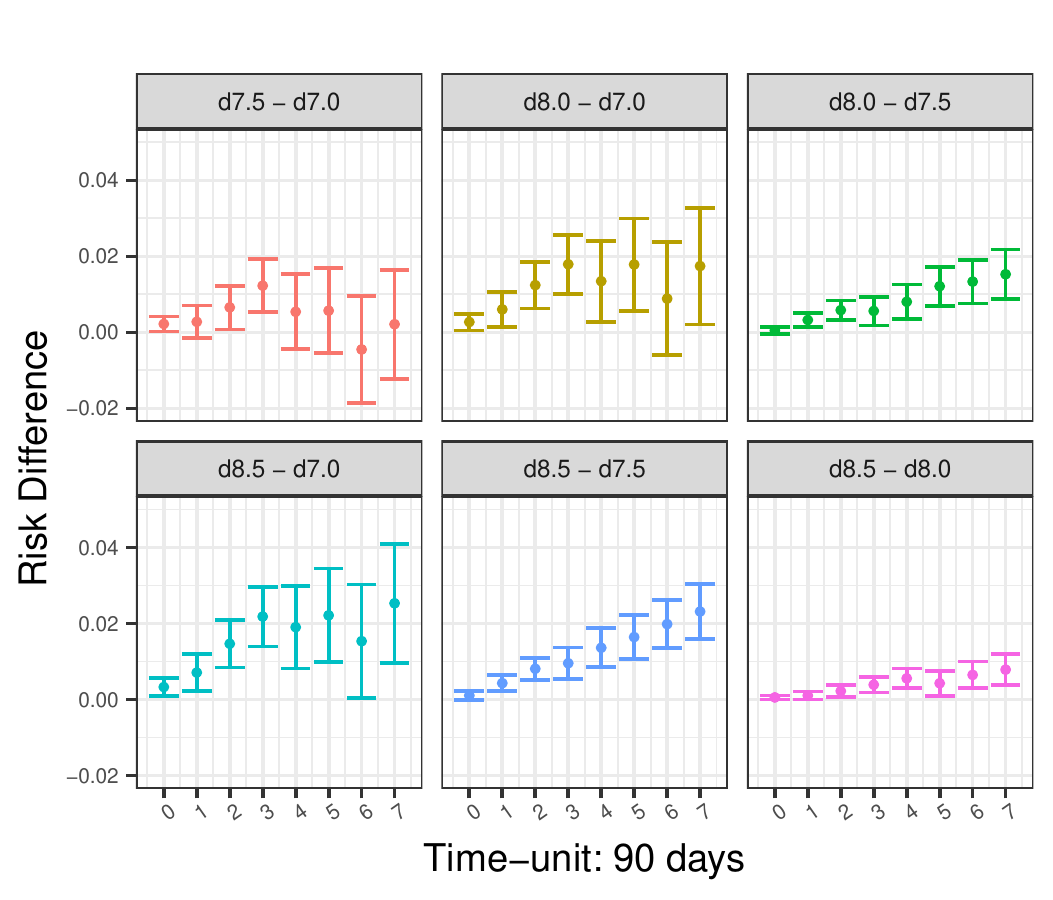}\includegraphics[scale=0.68]{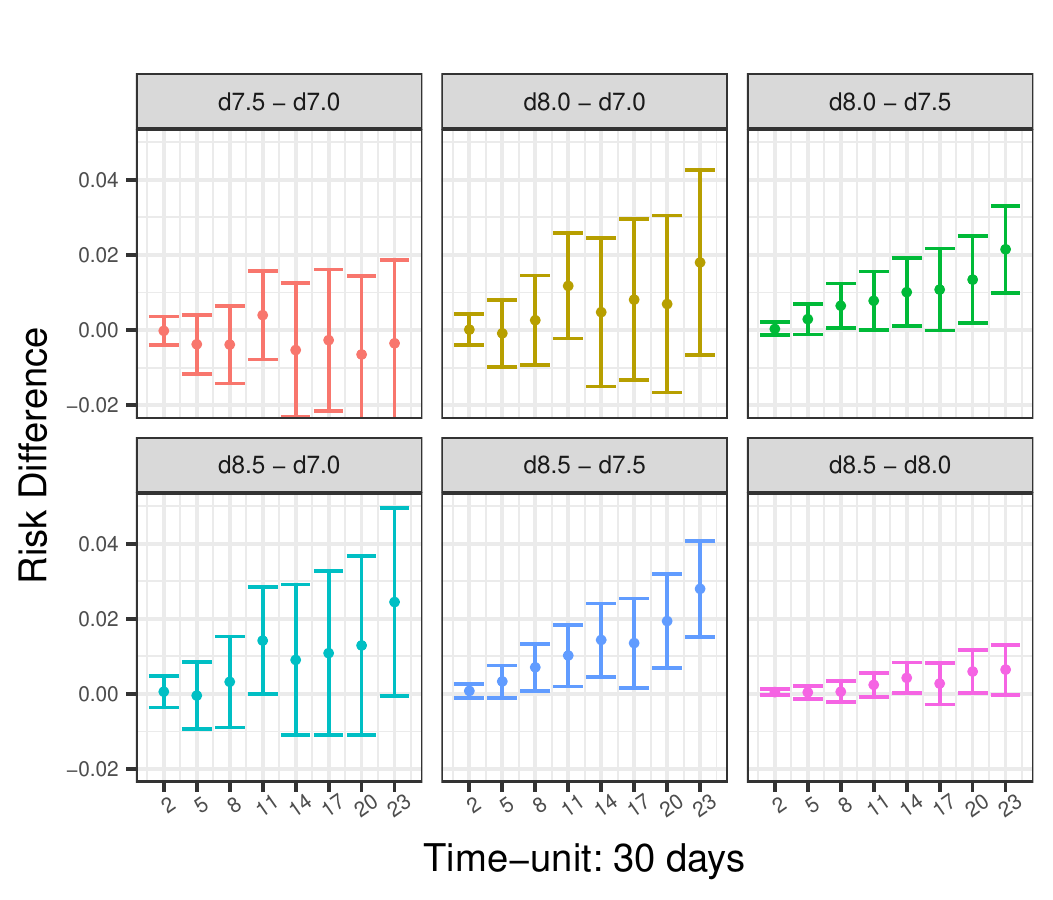}
\par\end{centering}

\centering{}\caption{TMLE estimates of cumulative RDs with point-wise 95\% CIs, contrasting
four dynamic interventions $(d_{7.0},d_{7.5},d_{8.0},d_{8.5})$ for 90
day (left panel) and 30 day (right panel) time-unit over two years of follow-up.\label{fig:TMLE-SL-90-30}}
\end{figure}

\begin{figure}
\begin{centering}
\includegraphics[scale=0.68]{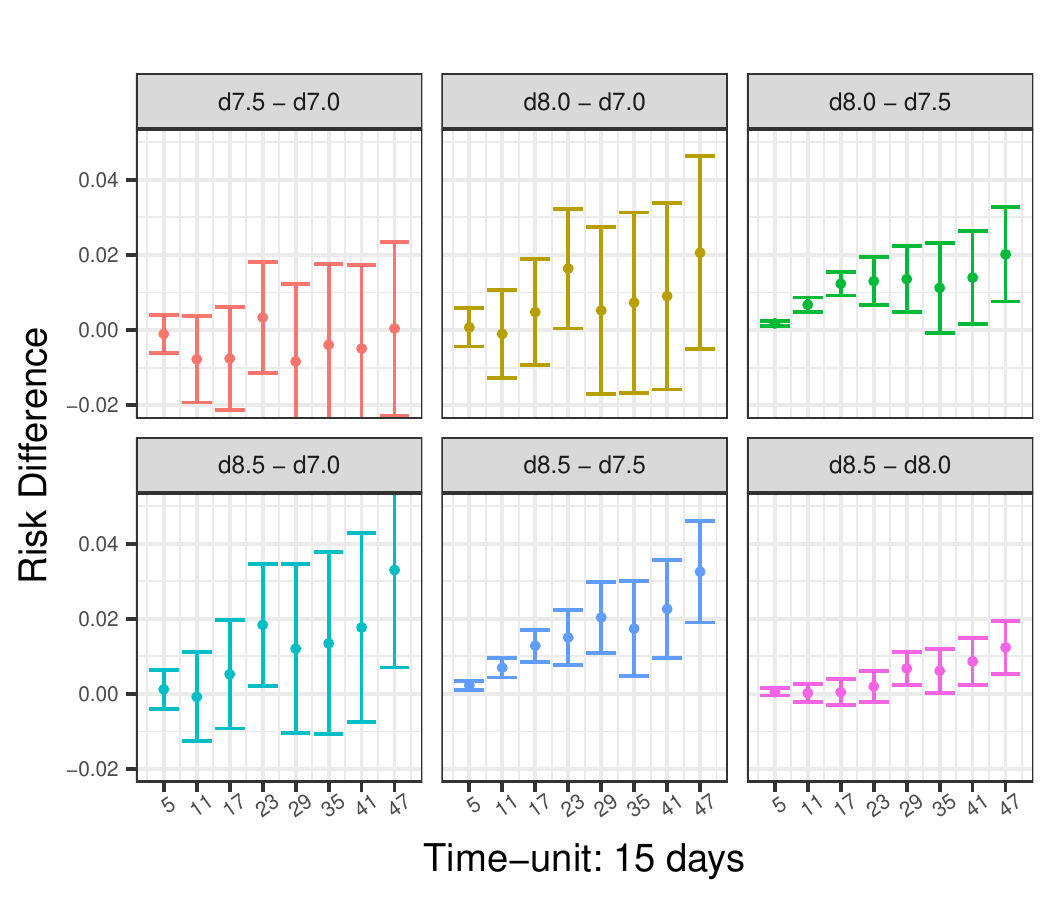}\includegraphics[scale=0.68]{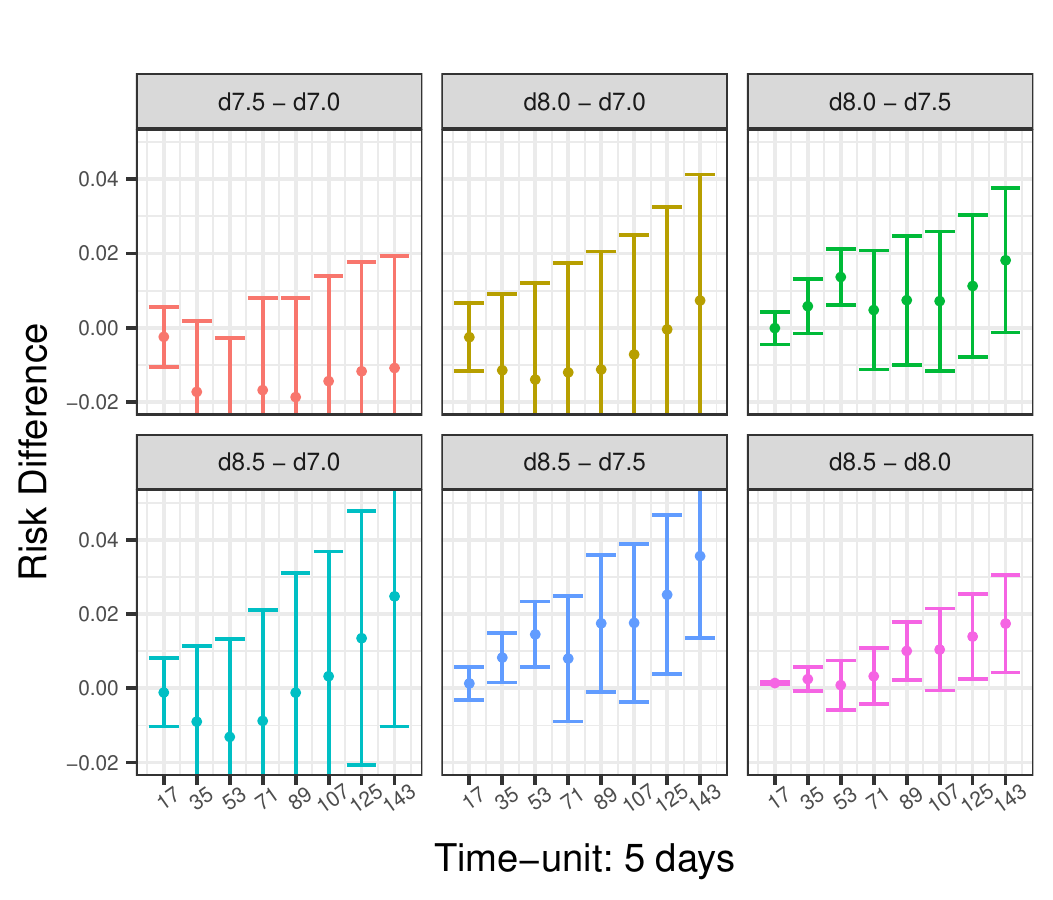}
\par\end{centering}

\centering{}\caption{TMLE estimates of cumulative RDs with point-wise 95\% CIs, contrasting
four dynamic interventions $(d_{7.0},d_{7.5},d_{8.0},d_{8.5})$ for 15
day (left panel) and 5 day (right panel) time-unit over two years of follow-up.\label{fig:TMLE-SL-15-5}}
\end{figure}

In this section, we demonstrate a possible application of our proposed
targeted learning software. We show that the $\mathtt{stremr}$ and $\mathtt{gridisl}$
R packages provide fast and scalable software for the analyses of high-dimensional
longitudinal data. We estimate the effects of four dynamic treatment regimes
on a time-to-event outcome using EHR data from the diabetes cohort study
described in Sections~\ref{sec:question} and \ref{sub:data-parameter}.
These analyses are based on large EHR cohort study, using four progressively
smaller time units (i.e., four nested discretizations of the same follow-up
data). For instance, the choice of the smaller time unit might more closely
approximate the original EHR daily event dates, as it is the case in our
application. We also evaluate the practical impact of these four progressively
smaller time units on inferences. Finally, we compare the long-format TMLE
results to those obtained from IPW estimator. All results are also compared
to prior published findings from alternate IPW and TMLE analyses.

The choice of time-unit of 90, 30, 15, and 5 days results in four analytic
datasets, each constructed by applying the SAS macro $\mathtt{\%\_MSMstructure}$
\citep{MSMmacro} to coarsen the original EHR data. The maximum follow-up
in each dataset is subsequently truncated to the first two years, i.e.,
8 quarters, 24 months, 48 15-day intervals, and 144 5-day intervals, respectively.
For each analytic dataset, we evaluate the counterfactual cumulative risks
at $t_{0}$ associated with four treatment intensification strategies,
with $t_{0}$ fixed to $8$ distinct time points. That is, for the 90-day
(resp. 30-day) analytic dataset, $\Psi^{\theta}(t_{0})(P)$ is estimated
for $t_{0}=0,1,\ldots,7$ (resp. $t_{0}=2,5,\ldots,23$) and $\theta=7,7.5,8,8.5$.
Similarly, for the 15-day (resp. 5-day) analytic dataset, $\Psi^{\theta}(t_{0})(P)$
is estimated for $t_{0}=5,10,\ldots,47$ (resp. $t_{0}=17,35,\ldots..,143)$
and $\theta=7,7.5,8,8.5$.

With each of the four analytic datasets and for each of the 32 target cumulative
risks, we evaluate the following two estimators: the stratified long-format
TMLE (Section~\ref{sub:lfTMLE}) and a bounded IPW estimator (based on
a saturated MSM for counterfactual hazards \citep{neugebauer2016}). The
TMLE and IPW estimators are implemented based on unstabilized and stabilized
IP weights truncated at 200 and 40, respectively \citep{cole2008}. Each
of the two estimators above uses two alternative strategies for nuisance
parameter estimation: a-priori specified logistic regression models (\emph{parametric
approach}) and discrete super learning with 10-fold cross-validation (\emph{data-adaptive
approach}).

\subsection{Nuisance parameter estimation approaches}

For the parametric approach, the estimators of each $Q_{k}$ and $g_{A(k)}$,
for $k=0,\ldots,t_{0}$, are based on separate logistic regression models
that include main terms for all baseline covariates $L(0)$, the exposures
$(A(k),A(k-1))$ and the most recent measurement of time-varying covariates
$L(k)$, i.e., the reduced dataset is defined with $f_{k}(L'(k))=(L(0),L(k),A(k-1))$.
This covariate selection approach results in approximately 150 predictors
for each regression model for $Q_{k}$ and $g_{A(k)}$, at each time point
$k$. In addition, estimation of $g$ relies on fitting separate logistic
models for treatment initiation and continuation. Furthermore, logistic
models for two types of right-censoring events (health plan disenrollment
and death) are fit separately, while it is assumed that right-censoring
due to end of the study is completely at random (i.e., we use an intercept-only
logistic regression). Each of the logistic models for estimating $g$ are
fit by pooling data over all time-points $k=0,\ldots t_{0}$. Furthermore,
the same sets of predictors that are used for estimation of $Q_{k}$ are
also included in estimation of $g$, in addition to a main term for the
value of time $k$. Finally, for $30$, $15$ and $5$ day time units,
these logistic models also include the indicators that the follow-up time
$k$ belongs to a particular two-month interval.

\begin{remark} The $\mathtt{\%\_MSMstructure}$ SAS macro \citep{MSMmacro}
implements automatic imputation of the missing covariates and creates indicators
of imputed values. The indicators of imputation are also included in each
$L(k)$. The documentation for the SAS macro provides a detailed description
of the implemented schemes for imputation. \end{remark}

For the data-adaptive estimation approach, each of the above-described
logistic model-based estimators is replaced with a distinct discrete super
learner. Each discrete super learner uses the same set of predictors that
are included in the corresponding model from the parametric approach. The
replication R code for the specification of each dSL is available from
the following github repository: \href{http://www.github.com/osofr/stremr.paper}{www.github.com/osofr/stremr.paper}.
In short, for estimation of each component of $g$, for 90, 30 and 15 day
time-unit, the dSL uses the following ensemble of 89 distributed (i.e.,
parallelized) estimators that are each indexed by a particular tuning parameter
choice: Random Forests (RFs) with $\mathtt{h2o}$ (8); Gradient Boosting
Machines (GBMs) with $\mathtt{h2o}$ (1); GBMs with $\mathtt{xgboost}$
(18); Generalized Linear Models (GLMs) with $\mathtt{h2o}$ (2); and regularized
GLMs with $\mathtt{h2o}$ (60). For 5 day time-unit, the dSL for each component
of $g$ uses the following smaller ensemble of 20 estimators: RFs with
$\mathtt{h2o}$ (1); GBMs with $\mathtt{h2o}$ (1); GBMs with $\mathtt{xgboost}$
(1); GLMs with $\mathtt{h2o}$ (2); and regularized GLMs with $\mathtt{h2o}$
(15). An abbreviated discussion of some of the tuning parameters that index
the ML algorithms considered is provided in Remark~\ref{remark:tuning}.
To obtain discrete super learning estimates for each component of $Q=(Q_{k}:k=0,\ldots,t_{0})$,
we rely solely on the candidate estimators available in the $\mathtt{xgboost}$
R package because of computational constraints. Specifically, the dSL ensemble
for each $Q_{k}$, for all four analytic datasets, is restricted to the
following 8 estimators: GBMs with $\mathtt{xgboost}$ (3); GLMs with $\mathtt{xgboost}$
(1); and regularized GLMs with $\mathtt{xgboost}$ (4).

\begin{remark} To achieve maximum computational efficiency with $\mathtt{stremr}$,
it is essential to be able to parallelize the estimation of $\Psi_{t_{0}}$
over multiple time points $t_{0}$. However, the estimators implemented
in the most recent $\mathtt{h2o}$ version 3.10.4.7 do not allow outside
parallelization for different values of $t_{0}$ (i.e., $\mathtt{h2o}$
does not allow fitting two distinct discrete super learners in parallel).
For this reason, the data-adaptive estimation of each component of $Q$
was performed solely with $\mathtt{xgboost}$ R package. \end{remark}

\begin{remark}\label{remark:tuning} We use the following parameters to
fine-tune the performance of GBMs and RFs in $\mathtt{h2o}$ and $\mathtt{xgboost}$
R packages: $\mathtt{ntrees}$ $(\mathtt{h2o})$ and $\mathtt{nrounds}$
$(\mathtt{xgboost})$, $\mathtt{max\_depth}$ $(\mathtt{h2o}$ and $\mathtt{xgboost})$,
$\mathtt{sample\_rate}$ $(\mathtt{h2o})$ and $\mathtt{subsample}$ $(\mathtt{xgboost})$,
$\mathtt{col\_sample\_rate\_per\_tree}$ $(\mathtt{h2o})$ and $\mathtt{colsample\_bytree}$
$(\mathtt{xgboost})$, $\mathtt{learn\_rate}$ $(\mathtt{h2o})$ and $\mathtt{learning\_rate}$
$(\mathtt{xgboost})$. Furthermore, $\mathtt{xgboost}$ provides additional
``shrinkage'' tuning parameters, $\mathtt{max\_delta\_step}$ and $\mathtt{lambda}$.
These parameters allow controlling the smoothness of the resulting fit,
with higher values generally resulting in a more conservative estimator
fit that might be less prone to overfitting. Furthermore, these two tuning
parameters can be useful to reduces the risk of spurious predicted probabilities
that are near 0 and 1 and thus might help obtain a more stable propensity
score fit of $g$. Finally, for regularized logistic regression with $\mathtt{h2o}$
R package, we use $\mathtt{lambda\_search}$ option for computing the regularization
path and finding the optimal regularization value $\lambda$ \citep{friedman2010regularization}.
Similarly, for regularized logistic regressions in $\mathtt{xgboost}$
we use a grid of candidate $\lambda$ values and select the optimal value
by minimizing the cross validation mean-squared error.\end{remark}

\subsection{Benchmarks\label{sec:benchmarks}}

Our benchmarks provide the running times for conducting the above described
analyses with long-format TMLE, using the four EHR datasets. We report
separate running times for estimation of the nuisance parameter $g$ and
the TMLE $\hat{Q}^{*}$. All analyses were implemented on Linux server
with 32 cores and 250GB of RAM. Whenever possible, the computation was
parallelized over the available cores. For instance, the data-adaptive
estimation of $g$ and $Q$ was parallelized by using the distributed machine
learning procedures implemented in $\mathtt{h2o}$ and $\mathtt{xgboost}$
R packages. Similarly, the estimation of TMLE survival at 8 different time-points
$t_{0}$ was parallelized by the $\mathtt{stremr}$ R package. The compute
times (in hours) are presented in Table~\ref{tab:Benchmarks}. These results
are based on the two estimation strategies for the nuisance parameters,
as described above. For example, the sum of dSL $\hat{g}$ and dSL $\hat{Q}^{*}$
for 5 day time-unit is the total time it takes to obtain the results for
all 4 survival curves at the bottom of the right-panel in Figure \ref{fig:surv-TMLE-SL}.
These results show that dSL is computationally costly, but the running
times do not preclude routine application of the $\mathtt{stremr}$ and
$\mathtt{gridisl}$ R packages in EHR-based datasets, even for studies
that use a small time-unit.

\subsection{Results}

The long-format TMLE survival estimates for data-adaptive estimation of
$(g,Q)$ for all four time-units are presented in Figure \ref{fig:surv-TMLE-SL}.
Additional results using 90 and 30-day time-unit are presented in Figure
\ref{fig:TMLE-SL-90-30}, the corresponding results for 15 and 5-day time-unit
are presented in Figure \ref{fig:TMLE-SL-15-5}. All plots in Figures \ref{fig:TMLE-SL-90-30}
and \ref{fig:TMLE-SL-15-5} present the TMLE point estimates of risk differences
(RDs) for any two of the four dynamic regimens. The estimates are plotted
for 8 different values of $t_{0}$, along with their corresponding 95\%
confidence intervals (CIs). Results from the long-format TMLE analyses
with parametric estimation of $(Q,g)$ are presented in Figures 1-3 of
the Web Supplement D. Results from the IPW for parametric and data-adaptive
estimation strategies are presented in Figures 4-9 of the Web Supplement
D. Finally, the distributions of the untruncated and unstabilized IP weight
estimates $\hat{wt}(t)$ obtained with the two estimation strategies for
$g$ are presented in Tables 1 and 2 of the Web Supplement D.

The top-left panel in Figure \ref{fig:surv-TMLE-SL} and top panel in Figure
\ref{fig:TMLE-SL-90-30} demonstrates that we have replicated the prior
TMLE results for the 90-day time-unit from \citet{neugebauer2014}. The
point estimates from all four analyses provide consistent evidence, suggesting
that earlier treatment intensification provides benefits in lowering the
long term cumulative risk of onset or progression of albuminuria. However,
the new results are inconclusive for the earliest treatment intensification
with dynamic regime $d_{7.0}$ due to increasing variability of the estimates
with progressively smaller time-unit. Moreover, this trend is also observed
for the other three dynamic rules, as the variance estimates increase substantially
with the smaller time-unit. Finally, the data-adaptive approaches clearly
produce tighter confidence intervals, than with the logistic regression
alone.

The distribution of the propensity score based weights for each time unit
of analyses is also reported as part of the same supplementary materials.
Finally, we conduct an alternative set of analyses by including a large
number of two-way interactions for estimation of each $Q_{k}$, for $k=0,\ldots,t_{0}$.
However, these analyses did not materially change our findings and the
results are thus omitted.

\section{Discussion\label{sec:discussion}}

In this work we've studied the impact of choosing a different time-unit
on inference for comparative effectiveness research in EHR data. Current
guidelines suggest choosing the time-unit of analysis by dividing the granular
(e.g., daily) subject-level follow-up into small (and equal) time interval.
Relying on overly discretized EHR data might invalidate the validity of
the analytic findings in a number of ways. For example, {na\"{\i}ve} discretization
might introduce measurement error in the true observed exposure, accidentally
reverse the actual time ordering of the events, and may result in failure
to adjust for all \emph{measured} time-varying confounding. Thus, choosing
a small time-unit is a natural way to reduce the reliance on ad-hoc data-coarsening
decisions. As we've shown in our data analysis, the choice of time-unit
may indeed impact the inference.

The size and dimensionality of the currently available granular EHR data
presents novel computational challenges for application of semi-parametric
estimation approaches, such as longitudinal TMLE and data-adaptive estimation
of nuisance parameters. In our example, the actual EHR data is generated
daily from patient's encounters with the healthcare system. To address
these challenges we have developed and applied the ``long-format TMLE''
-- a new algorithmic solution for the existing targeted learning methodology.
We also apply a data-adaptive approach of discrete super learning (dSL)
to estimation of the corresponding nuisance parameters, based on its novel
implementation in the $\mathtt{gridisl}$ R package. Our benchmarks show
that $\mathtt{stremr}$ and $\mathtt{gridisl}$ R packages can be routinely
applied to EHR-based datasets, even for studies that use a very small time-unit.

Our analyses demonstrate a substantial increase in the variance of the
estimates associated with the selection of the smaller time-unit. As a
possible explanation, it should be pointed out that the choice of the smaller
time-unit increases the total number of considered time-points and will
typically result in a larger set of time-varying covariates. As a result,
one would expect that the estimation problem becomes harder for the smaller
time-unit (in part, due to the growing dimensionality of the time-varying
covariate sets, and, in part, due to larger number of nuisance parameters
that need to be estimated). This can potentially lead to a larger variance
of the underlying estimates, as was observed in our applied study. It remains
to be seen if a single estimation procedure could leverage different levels
of discretization and the choice of the different time-unit within the
same dataset to provide a more precise estimate. However, we leave the
formal methodological analysis of this subject for future research.

Finally, we point out that the $\mathtt{stremr}$ R package implements
additional estimation procedures that are outside the scope of this paper,
e.g., long-format TMLE for stochastic interventions, handling problems
with multivariate and categorical exposures at each time-point, no-direct-effect-based
estimators \citep{neugebauer2017} of joint dynamic treatment and monitoring
interventions, iterative longitudinal TMLE \citep{vanderLaan:Gruber12a},
sequentially double robust procedures, such as the infinite-dimensional
TMLE \citep{luedtke2017}, and other procedures described in the package
documentation and the following github page: \href{http://www.github.com/osofr/stremr}{www.github.com/osofr/stremr}.

\section*{Acknowledgements}

The authors thank the following investigators from the HMO research network
for making data from their sites available to this study: Denise M. Boudreau
(Group Health), Connie Trinacty (Kaiser Permanente Hawaii), Gregory A.
Nichols (Kaiser Permanente Northwest), Marsha A. Raebel (Kaiser Permanente
Colorado), Kristi Reynolds (Kaiser Permanente Southern California), and
Patrick J. O\textquoteright Connor (HealthPartners). This study was supported
through a Patient-Centered Outcomes Research Institute (PCORI) Award (ME-1403-12506).
All statements in this report, including its findings and conclusions,
are solely those of the authors and do not necessarily represent the views
of the Patient-Centered Outcomes Research Institute (PCORI), its Board
of Governors or Methodology Committee. This work was also supported through
an NIH grant (R01 AI074345-07).

\clearpage

\section*{Web Appendix A. Representing the targeted parameter as a sequence of recursively
defined iterated conditional expectations}

In this section we show that our representation of the statistical target
parameter from the main text is indeed a valid mapping for the desired
statistical quantity. We will first generalize the framework presented
in main text to the case of arbitrary stochastic interventions. We will
then present the mapping of the statistical estimand for an arbitrary stochastic
intervention in terms of iterated conditional expectations. We will finish
the proof by showing that the dynamic intervention considered in main text
of the paper is just a special case of a stochastic intervention, making
the presented proof also valid for the statistical estimand considered
in the main text of the paper.

\subsection*{Observed data, likelihood and the statistical model}

Suppose we observe $n$ i.i.d. copies of a longitudinal data structure
\[
O=(L(0),A(0),\ldots,L(\tau),A(\tau),L(\tau+1)=Y)\sim P,
\]
where $A(t)$ denotes binary valued intervention node, $L(0)$ baseline
covariates, $L(t)$ is a time-dependent confounder realized after $A(t-1)$
and before $A(t)$, for $t=0,\ldots,\tau$ and $L(\tau+1)=Y$ is the final
(binary) outcome of interest. There are no restrictions on the dimension
and support of $L(t)$, $t=0,\ldots,\tau$ and we assume that the outcome
of interest $Y$ at $\tau+1$ is always observed (i.e., no right-censoring
censoring, the data structure in $O$ is never degenerate for any $t=\tau,\ldots,0$).
The density $p(o)$ of $O\sim P$ can be factorized according to the time-ordering
as 
\begin{eqnarray}
p(o) & = & \prod_{t=0}^{\tau+1}p(l(t)\mid\bar{l}(t-1),\bar{a}(t-1))\prod_{t=0}^{\tau}p(a(t)\mid\bar{l}(t),\bar{a}(t-1))\label{P}\\
 & = & \prod_{t=0}^{\tau+1}q_{t}(l(t)\mid\bar{l}(t-1),\bar{a}(t-1))\prod_{t=0}^{\tau}g_{t}(a(t)\mid\bar{l}(t),\bar{a}(t-1)),
\end{eqnarray}
for some realization $o$ of $O$. Recall that $\bar{l}(t)=(l(0),...,l(t))$
and $\bar{a}(t)=(a(0),...,a(t))$. Also recall that $((l(-1),a(-1))$ is
defined as an empty set. Note also that $q_{t}$ denotes the conditional
density of $L(t)$, given $\bar{L}(t-1),\bar{A}(t-1)$, while $Q_{t}$
denotes its conditional distribution. Similarly, $g_{t}$ denotes the conditional
density of $A(t)$, given $(\bar{L}(t),\bar{A}(t-1))$, while $G_{t}$
denotes its conditional distribution. We assume the densities $q_{t}$
are well-defined with respect to some dominating measures $\mu_{L(t)}$,
for $t=0,\ldots\tau+1$ and $g_{t}$ are well-defined with respect to some
dominating measures $\mu_{A(t)}$, for $t=0,\ldots,\tau$. Similarly, assume
the density $p$ of $O$ is a well-defined density with respect to the
product measure $\mu$. Let $q=(q_{t}:t=1,\ldots,\tau+1)$ and $g=(g_{t}:t=1,\ldots,\tau)$,
so that the distribution of $O$ is parameterized by $(q,g)$. Consider
a statistical model $\mathcal{M}$ for $P$ that possibly assumes knowledge
on $g$. If $\mathcal{Q}$ is the parameter set of all values for $q$
and $\mathcal{G}$ the parameter set of possible values of $g$, then this
statistical model can be represented as $\mathcal{M}=\{P_{q,g}=QG\::\:q\in\mathcal{Q},g\in\mathcal{G}\}$.
In this statistical model $q$ puts no restrictions on the conditional
distributions $Q_{t}$ of $L(t)$ given $(\bar{L}(t-1),\bar{A}(t-1))$,
for $t=0,\ldots,\tau+1$.

\subsection*{Stochastic interventions}

Define the intervention of interest by replacing the conditional distribution
$G$ with a new user-supplied intervention $G^{*}=\{G_{t}^{*}:t=0,\ldots,\tau\}$
that has a density $g^{*}=\{g_{t}^{*}:t=0,\ldots,\tau\}$, which we assume
is well-defined. Namely, $G^{*}$ is a multivariate conditional distribution
that encodes how each intervened exposure $A^{*}(t)$ is generated, conditional
on $(\bar{L}(t),\bar{A}(t-1))$, for $t=0,\ldots,\tau$. When $g_{t}^{*}$
is non-degenerate it is often referred to as a ``stochastic intervention''
\citep{dawid2010identifying,Didelezetal06}. Furthermore, any static or
dynamic intervention \citep{Gill:Robins01} on $A(t)$ can be formulated
in terms of a degenerate choice of $g_{t}^{*}$. Therefore, the stochastic
interventions are a natural generalization of static and dynamic interventions.
We make no further restrictions on $G_{t}^{*}$ beyond assuming that $A(t)$
and $A^{*}(t)$ belong to the same common space $\mathcal{A}=\{0,1\}$
for all $t=0,\ldots,\tau$.

\subsection*{G-computation formula and statistical parameter}

Define the post-intervention distribution $P_{q,g^{*}}$ by replacing the
factors $G$ in $P_{q,g}$ with a new user-supplied stochastic intervention
$G^{*}$, with its corresponding post-intervention density $p_{q,g^{*}}$
given by 
\begin{equation}
p_{q,g^{*}}(o)=\prod_{t=0}^{\tau+1}q_{t}(l(t)\mid\bar{l}(t-1),\bar{a}(t-1))\prod_{t=0}^{\tau}g_{t}^{*}(a(t)\mid\bar{l}(t),\bar{a}(t-1)).\label{eq:post.like}
\end{equation}
The distribution $P_{q,g^{*}}$ of $p_{q,g^{*}}$ is referred to as the
\emph{G-computation formula} for the post-intervention distribution of
$O$, under the stochastic intervention $G^{*}$ \citep{robins1986new,vdL2014nets}.
It can be used for identifying the counterfactual post-intervention distribution
defined by a non-parametric structural equation model \citep{vanderLaan:Gruber12a,vdL2014nets}.

Let $O^{*}$ denote a random variable with density $p_{q,g^{*}}$ (\ref{eq:post.like})
and distribution $P_{q,g^{*}}$ , defined as a function of the data distribution
$P$ of $O$:
\[
O^{*}=(L(0),A^{*}(0),\ldots,L^{*}(\tau),A^{*}(\tau),L^{*}(\tau+1)=Y^{*}).
\]
Consider the statistical mapping $\Psi(P_{q,g})$ defined as 
\begin{equation}
\Psi(P_{q,g})=E_{P_{q,g^{*}}}Y^{*}=\int_{o\in\mathcal{O}}ydP_{q,g^{*}}(o).\label{psi}
\end{equation}
Note that the above mapping is defined with respect to the post-intervention
distribution $P_{q,g^{*}}$, and hence, $E_{P_{q,g^{*}}}Y^{*}$ is entirely
a function of the observed data generating distribution $P$ and the known
stochastic intervention $G^{*}$. In other words, $E_{P_{q,g^{*}}}Y^{*}$
is identified by the observed data distribution $P$ and it depends on
$P$ through $Q=(Q_{t}:t=0,\ldots,\tau+1)$.

Under additional causal identifying assumptions of sequential randomization
and positivity, as stated in following section, the statistical parameter
$E_{P_{q,g^{*}}}Y^{*}$ can be interpreted as the mean causal effect of
the longitudinal stochastic intervention $G^{*}$ on $Y$. However, formal
demonstration of these identifiability results requires postulating a causal
non-parametric structural equation model (NPSEM), with counterfactual outcomes
explicitly defined and additional explicit assumptions encoded by a causal
diagram \citep{pearl1995,pearl2009causal}. The G-computation formula and
the post-intervention distribution $P_{q,g^{*}}$ play the key role in
establishing these identifiability results. We refer the interested reader
to \citep{vanderLaan:Gruber12a} for the example application of the G-computation
formula towards identifiability of the average causal effect of static
regimens in longitudinal data with time-varying confounding. Note that
the latter causal parameter corresponds with the choice of a degenerate
$G^{*}$ that puts mass one on a single vector $\bar{a}(\tau)$. We also
refer to \citep{vdL2014nets} for the application of the post-intervention
distribution $P_{q,g^{*}}$ towards identifying the average causal effects
of arbitrary stochastic interventions in network-dependent longitudinal
data.

\subsection*{Target parameter as a function of iterated conditional means}

By applying the Fubini's theorem and re-arranging the order of integration,
we can re-write the target parameter $E_{P_{q,g^{*}}}Y^{*}$ in terms of
the iterated integrals as follows, 
\begin{align}
E_{P_{q,g^{*}}}Y^{*} & =\int_{o\in\mathcal{O}}ydP_{q,g^{*}}(o)\nonumber \\
 & =\int_{\bar{l}(\tau),\bar{a}(\tau)}\left[\int_{y}ydQ_{\tau+1}(y|\bar{l}(\tau),\bar{a}(\tau))\right]\prod_{t=0}^{\tau}dG_{t}^{*}(o)\prod_{t=0}^{\tau}dQ_{t}(o)\nonumber \\
 & =\int_{\bar{l}(\tau),\bar{a}(\tau)}\left[\int_{y}ydQ_{\tau+1}(y|\bar{l}(\tau),\bar{a}(\tau))\right]\prod_{t=0}^{\tau}\left\{ dG_{t}^{*}(o)dQ_{t}(o)\right\} .\label{eq:innerE}
\end{align}
For conciseness and with some abuse of notation, we used $G_{t}^{*}(o)$
and $Q_{t}(o)$ for denoting $G_{t}(a(t)\mid\bar{l}(t),\bar{a}(t-1))$
and $Q_{t}(l(t)\mid\bar{l}(t-1),\bar{a}(t-1))$, respectively, for $t=0,\ldots\tau$.
We also assumed that $\mathcal{O}$ represents the set of all values of
the observed data $O$. Note that the inner-most integral in the last line
is the conditional expectation of $Y$ with respect to the conditional
distribution $Q_{\tau+1}$ of $L(\tau+1)$ given $(\bar{A}(\tau),\bar{L}(\tau))$,
namely,
\begin{align*}
 & \bar{Q}_{\tau+1,1}\left(\bar{a}(\tau),\bar{l}(\tau)\right)\\
 & \equiv E_{Q_{\tau+1}}\left[Y|\bar{A}(\tau)=\bar{a}(\tau),\bar{L}(\tau)=\bar{l}(\tau)\right]\\
 & =\int_{y\in\mathcal{Y}}ydQ_{\tau+1}(y|\bar{l}(\tau),\bar{a}(\tau)).
\end{align*}
By applying Fubini's theorem one more time, we integrate out $a(\tau)$
with respect to the conditional stochastic intervention $G_{\tau}^{*}(o)$
as follows,

\begin{align*}
E_{P_{q,g^{*}}}Y^{*} & =\int_{\bar{l}(\tau),\bar{a}(\tau)}\left[\bar{Q}_{\tau+1,1}(\bar{a}(\tau),\bar{l}(\tau))\right]\prod_{t=0}^{\tau}\left\{ dG_{t}^{*}(o)dQ_{t}(o)\right\} \\
 & =\int_{\bar{l}(\tau),\bar{a}(\tau-1)}\left\{ \int_{a(\tau)}\left[\bar{Q}_{\tau+1,1}(\bar{a}(\tau),\bar{l}(\tau))\right]dG_{\tau}^{*}(o)\right\} dQ_{\tau}(o)\prod_{t=0}^{\tau-1}\left\{ dG_{t}^{*}(o)dQ_{t}(o)\right\} ,
\end{align*}
where we also note that the inner-most integral defines the following conditional
expectation,
\begin{align*}
 & \bar{Q}_{\tau+1}\left(\bar{a}(\tau-1),\bar{l}(\tau)\right)\\
 & \equiv E_{G_{\tau}^{*}}\left[\bar{Q}_{\tau+1,1}\left(\bar{A}(\tau),\bar{L}(\tau)\right)\left|\bar{A}(\tau-1)=\bar{a}(\tau-1),\bar{L}(\tau)=\bar{l}(\tau)\right.\right].
\end{align*}

We have now demonstrated the first two steps of the algorithm that represents
the integral $E_{p_{q,g^{*}}}Y^{*}$ in terms of the iterated conditional
expectations. The rest of the integration process proceed in a similar
manner, with the integration order iterated with respect to $(q_{t},g_{t-1}^{*})$,
for $t=\tau,\ldots,0$, moving backwards in time until we reach the final
expectation over the marginal distribution $Q_{0}$ of $L(0)$.

For notation convenience, let $\bar{Q}_{\tau+1,1}\equiv\bar{Q}_{\tau+1,1}(\bar{A}(\tau),\bar{L}(\tau))$
and $\bar{Q}_{\tau+1}\equiv\bar{Q}_{\tau+1}(\bar{A}(\tau-1),\bar{L}(\tau))$.
The full steps that represent $E_{p_{q,g^{*}}}Y^{*}$ as iterated conditional
expectations are as follows
\[
\begin{array}{l}
\mbox{Iterate, \ensuremath{t=\tau,\ldots,0}}\\
\bar{Q}_{t+1,1}=E_{q_{t+1}}(\bar{Q}_{t+1}\mid\bar{A}(t),\bar{L}(t))\\
\bar{Q}_{t+1}=E_{g_{t}^{*}}(\bar{Q}_{t+1,1}\mid\bar{A}(t-1),\bar{L}(t))\\
\ldots\\
\bar{Q}_{t=0}=E_{L(0)}\bar{Q}_{1}\\
=E_{p_{q,g^{*}}}\bar{Y}^{*}
\end{array}
\]

Note that this representation allows the effective evaluation of $E_{p_{q,g^{*}}}Y^{*}$
by first evaluating a conditional expectation with respect to the conditional
distribution of $L(\tau+1)$, then the conditional mean of the previous
conditional expectation with respect to the conditional distribution of
$A^{*}(\tau)$, and iterating this process of taking a conditional expectation
with respect to $L(t)$ and $A^{*}(t-1)$ until we end up with a conditional
expectation over $A^{*}(0)$, given $L(0)$, and finally we take the marginal
expectation with respect to the distribution of $L(0)$.

\subsection*{Applying the iterative representation of the estimand for stochastic intervention
to fixed dynamic intervention}

Define $g_{t}^{*}$ for $t=0,\ldots,\tau$ so that they define our dynamic
intervention of interest. In particular, let $g_{t}^{*}$ be a degenerate
distribution that puts mass one on a single value of $a(t)$ and that is
equal to 
\[
g_{t}^{*}(a(t)|\bar{l}(t),\bar{a}(t-1))=I(a(t)=a^{\theta}(t)),
\]
for $t=0,\ldots\tau$. Then the above representation of $E_{P_{q,g^{*}}}\bar{Y}^{*}$
in terms of the iterated conditional expectations is equivalent to the
mapping $\Psi^{\theta}(\tau)$ as presented in the main text of the paper.
To see this, note that the above iterated integration steps with respect
to the stochastic intervention $g_{t}^{*}$ simplify to 
\begin{align*}
 & \bar{Q}_{t+1}\left(\bar{A}(t-1),\bar{L}(t)\right)\\
 & =E_{G_{t}^{*}}\left[\bar{Q}_{t+1,1}\left(\bar{A}(t),\bar{L}(t)\right)\left|\bar{A}(t-1),\bar{L}(t)\right.\right]\\
 & =\bar{Q}_{t+1,1}\left(\bar{A}(t),\bar{L}(t)\right)I\left(A(t)=A^{*}(t)\right)\\
 & =E_{Q_{t+1}}(\bar{Q}_{t+1}\mid A(t)=A^{*}(t),\bar{A}(t-1),\bar{L}(t)),
\end{align*}
for $t=0,\ldots,\tau$. By plugging this result back into the above described
iterated means mapping of $E_{P_{q,g^{*}}}\bar{Y}^{*}$, we obtain exactly
the same sequential G-computation mapping as the one presented in the main
text of this paper for the parameter $\Psi^{\theta}(\tau)$. This finishes
the proof, since it shows that indeed our statistical parameter representation
in the main text is valid, i.e., it produces the desired statistical estimand
identified by the post-intervention G-computation formula.

\section*{Web Appendix B. Causal parameter and causal identifying assumptions\label{subsec:Causal-assump}}

Let $g^{*}=\{g_{t}^{*}:t=0,\ldots,\tau\}$ denote the stochastic intervention
of interest which determines the random assignment of the observed treatment
nodes $\bar{A}=(A(0),\ldots,A(\tau))$. Let $Y_{g^{*}}$ denote the patient's
potential outcome at time $\tau+1$ had the patient been treated according
to the randomly drawn treatment strategy $g^{*}$. Similarly, let $L_{g^{*}}(t)$
denote the counterfactual values of the patient's time-varying or baseline
covariates at time $t=0,\ldots,\tau$, under intervention $g^{*}$. Finally,
let $EY_{g^{*}}$ denote the causal parameter of interest, defined as the
mean causal effect of intervention $g^{*}$ on the outcome.

The causal validity of the statistical estimand presented above rests on
the following two untestable identifying assumptions:

\begin{assume}[Sequential Randomization Assumption (SRA)] For each $t=0,\ldots,\tau$,
assume that $A(t)$ is conditionally independent of $(L_{g^{*}}(t+1),\ldots,L_{g^{*}}(\tau),Y_{g^{*}},)$
given the observed past $(\bar{L}(t),\bar{A}(t-1))$. \end{assume}

\begin{assume}[Positivity Assumption (PA)] For $k=0,\ldots,\tau$, assume
\[
\sup_{o\in\mathcal{O}}\dfrac{\prod_{t=0}^{k}g_{t}^{*}(a(t)|\bar{l}(t),\bar{a}(t-1))}{\prod_{k=0}^{k}g_{t}^{*}(a(t)|\bar{l}(t),\bar{a}(t-1))}<\infty\text{ }\text{ }P_{0}-a.e.,
\]
where $\mathcal{O}$ is the support of $O=(L(0),A(0),\ldots,L(\tau),A(\tau),L(\tau+1)=Y)\sim P_{0}$.\end{assume}

One can obtain the causal identifiability results for dynamic intervention
$(A_{i}^{\theta}(0),\ldots,A_{i}^{\theta}(\tau))$ by simply letting $g_{t}^{*}$
be a degenerate distribution that puts mass one on a single value of $a(t)$
and setting it equal to 
\[
g_{t}^{*}(a(t)|\bar{l}(t),\bar{a}(t-1))=I(a(t)=a^{\theta}(t)),
\]
for $t=0,\ldots\tau$. 

\section*{Web Appendix C. Summary of practical implementation of TMLE for fixed
dynamic rule}

We describe the long-format TMLE algorithm for estimating parameter $\Psi^{\theta}(t_{0})(P)$
indexed by the fixed dynamic regimen $\bar{d}_{\theta}=(d_{\theta,0},...,d_{\theta,t_{0}})$.
For notational convenience we let $t_{0}=1$. To define TMLE we need to
use the efficient influence curve (EIC) of the statistical target parameter
$\Psi^{\theta}(t_{0})$ at $P$, which is given by: 
\[
D^{*}(\theta,t_{0})(P)=D^{0,*}(\theta,t_{0})(P)+D^{1,*}(\theta,t_{0})(P)+D^{2,*}(\theta,t_{0})(P),
\]
 where 
\begin{eqnarray*}
D^{0,*}(\theta,t_{0})(P) & = & \left(Q_{0}-\Psi^{\theta}(t_{0})\right)\\
D^{1,*}(\theta,t_{0})(P) & = & \frac{I(A(0)=A^{\theta}(0))}{g_{A(0)}(O)}(Q_{1}-Q_{0})\\
D^{2,*}(\theta,t_{0})(P) & = & \frac{I(\bar{A}(1)=\bar{A}^{\theta}(1))}{g_{A(0)}(O)g_{A(1)}(O)}\left(Y(t_{0})-Q_{1}\right).
\end{eqnarray*}

The TMLE algorithm described below involves application of the sequential
G-computation formula from \citet{Bang:Robins05}. In this implementation
of TMLE one carries out the TMLE update step by fitting a separate $\varepsilon^{j}$
for updating each $Q_{j}$, for $j=1,0$, and sequentially carrying out
these updates starting with $Q_{1}$ and going backwards. In addition,
it involves first targeting the regression before defining it as outcome
for the next regression backwards in time. For our particular example with
two time-points, this entails obtaining an estimate $\hat{Q}_{1}$ of $Q_{1}$,
running the first TMLE update on $\hat{Q}_{1}$ to obtain a targeted estimate
$\hat{Q}_{1}^{*}$ . This is followed by using the estimate $\hat{Q}_{1}^{*}$
to obtain an initial estimate $\hat{Q}_{0}$ of $Q_{0}$ and running the
second TMLE update on $\hat{Q}_{0}$ to obtain a targeted estimate $\hat{Q}_{0}^{*}$.
Finally, TMLE estimate $\hat{\Psi}^{\theta}(t_{0})$ of $\Psi^{\theta}(t_{0})=E_{P}(Q_{0}(A^{\theta}(0),L'(0)))$
can be obtained from $\hat{Q}_{0}^{*}$ as $\hat{\Psi}^{\theta}(t_{0})=\frac{1}{n}\sum_{i=1}^{n}\hat{Q}_{0}^{*}(A_{i}^{\theta}(0),f_{0}(L'_{i}(0)))$.

In practice, the initial estimate $\hat{Q}_{1}$ is obtained by first regressing
the outcomes $Y_{i}(1)$ against $(A_{i}(1),L'_{i}(1))$, among the subjects
$i\in\{1,\ldots,n\}$ that at $t=1$: $a)$ were at risk of experiencing
the event of interest (i.e., $Y_{i}(t-1)=0$); $b)$ were uncensored at
$t$ (i.e., $A_{i}^{C}(t)=0$); and $c)$ (optionally) had their observed
exposure indicator at time $t$ match the values allocated by their dynamic
treatment rule $\bar{d}_{\theta}$ (i.e., $A_{i}^{T}(t)=A_{i}^{\theta,T}$).
More generally, fitting $Q_{1}$ can rely on data-adaptive techniques based
on the following log-likelihood loss function:
\[
L_{1}(Q_{1})=-\left\{ Y(1)\log Q_{1}+(1-Y(1))\log\left(1-Q_{1}\right)\right\} .
\]
The resulting model fit produces a mapping $(a(1),l'(1))\rightarrow E_{n}(Y(1)|a(1),l'(1))$
that can be now used to obtain an estimate $\hat{Q}_{1}$ of $Q_{1}$.
That is, the prediction $\hat{Q}_{1}$ is obtained as $E_{n}(Y_{i}(1)|A_{i}^{\theta}(1),L'_{i}(1))$,
for subjects $i$ such that $\tilde{T}_{i}\geq1$. However, for all subjects
such that $\tilde{T}_{i}=0$ and $A_{i}^{C}(0)=0$, $\hat{Q}_{1}$ is set
to $Y_{i}(0)$. Note that the later prediction $\hat{Q}_{1}$ is extrapolated
to all subjects who were also right-censored at $t=1$. The first TMLE
update modifies the initial estimate $\hat{Q}_{1}$ with its targeted version
$\hat{Q}_{1}^{*}$. This update utilizes the estimates $\hat{g}_{A(j)}$
of the joint exposure and censoring mechanism $g_{A(j)}$ for time-points
$j=0,1$. Furthermore, this update will be based on the least favorable
\emph{univariate} submodel (with respect to the target parameter $\psi_{0}$)
$\{\hat{Q}_{1}(\varepsilon^{1}):\varepsilon^{1}\}$ through a current fit
$\hat{Q}_{1}$ at $\varepsilon^{1}=0$, were the estimate $\hat{\varepsilon}^{1}$
of $\varepsilon^{1}$ is obtained with standard MLE. In practice, we define
this parametric submodel through $\hat{Q}_{1}$ as $\mbox{logit}\hat{Q}_{1}(\varepsilon^{1})=\mbox{logit}\hat{Q}_{1}+\varepsilon^{1}$
and we use the following weighted loss function function for fitting $\varepsilon^{1}$:
\[
L(Q_{1}(\varepsilon^{1}))\equiv H_{1}(\hat{g})L_{1}(Q_{1}(\varepsilon^{1})),
\]
where 
\[
H_{1}(\hat{g})=\dfrac{I(\bar{A}(1)=\bar{A}^{\theta}(1))}{\prod_{j=0}^{1}\hat{g}_{A(j)}(O)}
\]
and the fit of $\varepsilon^{1}$ defined by $\hat{\varepsilon}^{1}=\arg\min_{\varepsilon^{1}}L(Q_{1}(\varepsilon^{1}))$.
Thus, the estimate $\hat{\varepsilon}^{1}$ of $\varepsilon^{1}$ can be
obtained by simply running the intercept-only weighted logistic regression
using the sample of observations that were used for fitting $\hat{Q}_{1}$,
using the outcome $Y(1)$, intercept $\varepsilon^{1}$, the offset $\mbox{logit}\hat{Q}_{1}$
and the predicted weights $H_{1}(\hat{g})$. The fitted intercept is the
maximum likelihood fit $\hat{\varepsilon}^{1}$ for $\varepsilon^{1}$,
yielding the model update which can be evaluated for any fixed $(a,w)$,
by first computing the initial model prediction $\hat{Q}_{1}(a,w)$ and
then evaluating the model update $\hat{Q}_{1}(\hat{\varepsilon}^{1})$.
This now constitutes the first TMLE step, which we define as 
\[
\hat{Q}_{1}^{*}=\mbox{expit}\left(\mbox{logit}\hat{Q}_{1}+\hat{\varepsilon}^{1}\right),
\]
for subjects $i$ such that $\tilde{T}_{i}\geq1$.

The estimator $\hat{Q}_{0}$ of $Q_{0}$ is then obtained by selecting
subjects $i\in\{1,\ldots,n\}$ who were uncensored at $t=0$ (i.e., $A_{i}^{C}(t)=0$)
and (optionally) had their observed exposure matching the values allocated
by their dynamic treatment rule at $t=0$ (i.e., $A_{i}^{T}(t)=A_{i}^{\theta,T}(t)$).
The predicted outcomes $\hat{Q}_{1}^{*}(A_{i}^{\theta}(1),L'_{i}(1))$
are then regressed against $(A_{i}(0),L'_{i}(0))$ to obtain an estimate
$\hat{Q}_{0}$: $(a(0),l'(0))\rightarrow E_{n}(\hat{Q}_{1}^{*}|a(0),l'(0))$.
More generally, one can consider the following quasi-log-likelihood loss
functions for $Q_{0}$:

\[
L_{0,\hat{Q}_{1}^{*}}(Q_{0})=-\left\{ \hat{Q}_{1}^{*}\log Q_{0}+(1-\hat{Q}_{1}^{*})\log\left(1-Q_{0}\right)\right\} 
\]
and use data-adaptive techniques to obtain an estimate of $Q_{0}$. Finally,
$\hat{Q}_{0}$ is obtained from this regression function fit by evaluating
$E_{n}(\hat{Q}_{1}^{*}|A^{\theta}(0),L'(0))$ for $i=1,\ldots,n$, which
yields $n$ initial predictions $\hat{Q}_{0}$ of $Q_{0}$. Similar to
the update for $\hat{Q}_{1}$, the updated estimate $\hat{Q}_{0}^{*}$
of $\hat{Q}_{0}$ is obtained from the following least favorable \emph{univariate}
parametric submodel $\left\{ \hat{Q}_{0}(\varepsilon^{0}):\varepsilon^{0}\right\} $
through the current fit $\hat{Q}_{0}$:
\[
\mbox{logit}\hat{Q}_{0}(\varepsilon^{0})=\mbox{logit}\hat{Q}_{0}+\varepsilon^{0}
\]
and using the following weighted loss function for $\hat{Q}_{0}(\varepsilon^{0})$:
\[
L_{\hat{Q}_{1}^{*}}(Q_{0}(\varepsilon^{0}))\equiv H_{0}(\hat{g})L_{0,\hat{Q}_{1}^{*}}(Q_{0}(\varepsilon^{0})),
\]
where
\[
H_{0}(\hat{g})=\dfrac{I(A(0)=A^{\theta}(0))}{\hat{g}_{A(0)}(O)}.
\]

The corresponding TMLE $\hat{\Psi}^{\theta}(t_{0})$ of $\Psi^{\theta}(t_{0})$
is given by the following substitution estimator $\hat{\Psi}^{\theta}(t_{0})=\frac{1}{n}\sum_{i=1}^{n}\hat{Q}_{0}^{*}(A_{i}^{\theta}(0),L'_{i}(0))$.
The above loss functions and the corresponding least-favorable fluctuation
submodels imply that the TMLE $\hat{\Psi}^{\theta}(t_{0})$ solves the
the empirical score equation given by the efficient influence curve $D^{*}(\theta,t_{0})$.
That is, $\hat{\Psi}^{\theta}(t_{0})$ solves the estimating equation given
by $\frac{1}{n}\sum_{i=1}^{n}D^{*}(\theta,t_{0})(\hat{Q}_{1}^{*},\hat{Q}_{2}^{*},\hat{g})=0$,
implying that $\hat{\Psi}^{\theta}(t_{0})$ also inherits the double robustness
property of this efficient influence curve. Thus, the TMLE yields a substitution
estimator that empirically solves the estimating equation corresponding
to the efficient influence curve.

\section*{Web Appendix D. Regularity conditions and inference}

\subsection*{Regularity conditions}

The following theorem states the regularity conditions for asymptotic normality
of the TMLE $\hat{\Psi}^{\theta}(t_{0})$. In this discussion we limit
ourselves to providing the regularity conditions in a more limited setting
when both nuisance parameters, $Q$ and $g$, are both assumed to converge
to the truth ``fast enough'' rates, as clarified in conditions below.
However, when some or all estimates in $\hat{Q}=(\hat{Q}_{1},\ldots,\hat{Q}_{t_{0}})$
are incorrect, the asymptotic normality may still hold, e.g., when nuisance
parameters in $g$ converge to the truth at parametric rates. For a more
general discussion of such cases and the corresponding technical conditions
that guarantee the asymptotic normality of the TMLE we refer to \citet[Appendix 18]{vanderLaan:Rose11}
and \citet{van2015targeted}.

For a real-valued function $w\mapsto f(w)$, let the $L^{2}(P)$-norm of
$f(w)$ be denoted by $\norm{f}\equiv E[f(\mathbf{W})^{2}]^{1/2}$. Define
$\mathcal{F}$ and $\mathcal{G}$ as the classes of possible functions
that can be used for estimating $Q$ and $g$, respectively. The first
assumption below states the empirical process conditions which ensure that
the estimators of $Q$ and $g$ are well-behaved with probability approaching
one \citep{vanderVaart:Wellner96}. For a $K$-dimensional vector of functions
$\mathbf{f}=(f_{1},\ldots,f_{K})$ we also define $||\mathbf{f}||\equiv\max_{k\in\{1,\ldots,K\}}\left\{ ||f_{k}||\right\} $.
Finally, we let $Pf$ denote an expectation $E_{P}f(O)$ for any function
$f$ of $O$.

\begin{theorem}

If the below conditions hold then the TMLE $\hat{\Psi}^{\theta}(t_{0})$
is asymptotically linearly estimator of $\Psi^{\theta}(t_{0})$ at true
$P$ (true distribution of observed data), with the influence curve $D^{*}(\theta,t_{0})(Q,g)$
as defined in the previous section.

\end{theorem}
\begin{description}
\item [{Donsker class:}] Assume that $\{D^{*}(Q,g):Q,g\}$ is $P$-Donsker
class, for $Q\in\mathcal{F}$ and $g\in{\cal G}$. Assume that $g$ belongs
to a fixed class $\mathcal{G}$ with probability approaching one. 
\item [{Universal bound:}] Assume $\sup_{f\in{\cal F},O}\mid f\mid(O)<\infty$,
where the supremum of $O$ is over a set that contains $O$ with probability
one. Note that this condition will be typically satisfied by the above
Donsker class condition.
\item [{Positivity:}] Assume $0<\min_{o\in\mathcal{O}}\left\{ \prod_{j=0}^{t}g_{A(t)}(o)\right\} $,
for all $t$ and $\mathcal{O}$ representing the support of the random
variable $O$.
\item [{Consistent estimation of $D^*$:}] $P\left(D^{*}(\hat{Q}^{*},\hat{g})-D^{*}(Q,g)\right)^{2}\rightarrow0$
in probability, as $n\rightarrow\infty$.
\item [{Rate of the second order term:}]  Define the following second-order
term 
\[
R_{n}(\hat{Q}^{*},\hat{g},Q,g)\equiv P\left\{ D^{*}(\hat{Q}^{*},\hat{g})-D^{*}(\hat{Q}^{*},g)\right\} -P\left\{ D^{*}(Q,\hat{g})-D^{*}(Q,g)\right\} .
\]
Assume that $R_{n}(\hat{Q}^{*},\hat{g},Q,g)=o_{P}(1/\sqrt{n})$. 
\end{description}
Note that the above two conditions (rate and consistency) will be trivially
satisfied if one assumes the following \emph{stronger} condition:
\begin{description}
\item [{Consistency and rates for estimators of nuisance parameters:}] Assume
that $\norm{\hat{Q}-Q}\norm{\hat{g}-g}=o_{P}\left(n^{-1/2}\right)$, where
the norm $\norm{\cdot}$ for the $k$-dimensional function $f$ is defined
above. Note that these rates are achievable if these estimates $\hat{Q}$
and $\hat{g}$ are based on the corresponding correctly specified classes
$\mathcal{F}$ and $\mathcal{G}$.
\end{description}

\subsection*{Inference}

Estimation of the SEs for RDs can be based on the estimates of the efficient
influence curve (EIC) $D^{*}(P)$, where the EIC for our parameter of interest
is defined in Web Supplement B. The inference for the parameter $\Psi^{\theta}(t_{0})(P)$
can be based on the plug-in variance estimate $\hat{D}^{*}(\theta,t_{0})=D^{*}(\theta,t_{0})(\hat{Q}^{*},\hat{g})$
of the EIC $D^{*}(\theta,t_{0})(Q,g)$ defined in the Web Supplement B.
That is, the \emph{asymptotic} variance estimate of $\hat{\Psi}^{\theta}(t_{0})$
can be estimated as $\hat{\sigma}^{2}(\theta,t_{0})=\frac{1}{n}\sum_{i=1}^{n}\left[\hat{D}^{*}(\theta,t_{0})(O_{i})\right]^{2}$.
Furthermore, from the delta method, we know that the EIC for the risk difference
$\Psi^{\theta_{1}}(t_{0})-\Psi^{\theta_{2}}(t_{0})$ and single observation
$O_{i}$ is given by $D^{RD}(\theta_{1},\theta_{2},t_{0})(O_{i})=\left(D^{*}(\theta_{1},t_{0})-D^{*}(\theta_{2},t_{0})\right)(O_{i})$.
Thus, the asymptotic variance of the TMLE RD $\Psi_{n}^{\theta_{1}}(t_{0})-\Psi_{n}^{\theta_{2}}(t_{0})$
can be also estimated via the following plug-in estimator 
\[
\hat{\sigma}^{2}(\theta_{1},\theta_{2},t_{0})=\frac{1}{n}\sum_{i=1}^{n}\left[\left(\hat{D}^{*}(\theta_{1},t_{0})-\hat{D}^{*}(\theta_{2},t_{0})\right)(O_{i})\right]^{2}.
\]
Furthermore, Wald-type confidence intervals (CIs) can be now easily obtained
from these variance estimates, as described in detail in \citet{neugebauer2014}.

\clearpage

\section*{Web Appendix E. Additional analyses}

\subsection*{Summary of the IP-weights}

Note that all IP-weights are unstabilized.

\subsubsection*{Summary of the IP-weights for the data-adaptive approach}

\begin{table}[h]
{\tiny
\caption{Distribution of the IP-weights for data-adaptive modeling approach for dynamic intervention $d_{7.0}$ and 90-day time-unit.\label{res}}
\begin{center}
\begin{tabular}{lllll}
\hline\hline
\multicolumn{1}{c}{Stabilized IPAW}&\multicolumn{1}{c}{Frequency}&\multicolumn{1}{c}{\%}&\multicolumn{1}{c}{Cumulative Frequency}&\multicolumn{1}{c}{Cumulative \%}\tabularnewline
\hline
\textless 0&     0& 0.00&     0&  0.00\tabularnewline
$[$0, 0.5$[$&326730&96.44&326730& 96.44\tabularnewline
$[$0.5, 1$[$&     0& 0.00&326730& 96.44\tabularnewline
$[$1, 10$[$&  5381& 1.59&332111& 98.02\tabularnewline
$[$10, 20$[$&  3213& 0.95&335324& 98.97\tabularnewline
$[$20, 30$[$&  1494& 0.44&336818& 99.41\tabularnewline
$[$30, 40$[$&   754& 0.22&337572& 99.64\tabularnewline
$[$40, 50$[$&   456& 0.13&338028& 99.77\tabularnewline
$[$50, 100$[$&   691& 0.20&338719& 99.97\tabularnewline
$[$100, 150$[$&    73& 0.02&338792&100.00\tabularnewline
$\geq$ 150&    14& 0.00&338806&100.00\tabularnewline
\hline
\end{tabular}\end{center}}

\end{table}

\begin{table}[h]
{\tiny
\caption{Distribution of the IP-weights for data-adaptive modeling approach for dynamic intervention $d_{7.0}$ and 30-day time-unit.\label{res}}
\begin{center}
\begin{tabular}{lllll}
\hline\hline
\multicolumn{1}{c}{Stabilized IPAW}&\multicolumn{1}{c}{Frequency}&\multicolumn{1}{c}{\%}&\multicolumn{1}{c}{Cumulative Frequency}&\multicolumn{1}{c}{Cumulative \%}\tabularnewline
\hline
\textless 0&     0& 0.00&     0&  0.00\tabularnewline
$[$0, 0.5$[$&974355&97.89&974355& 97.89\tabularnewline
$[$0.5, 1$[$&     0& 0.00&974355& 97.89\tabularnewline
$[$1, 10$[$&  4531& 0.46&978886& 98.34\tabularnewline
$[$10, 20$[$&  3690& 0.37&982576& 98.72\tabularnewline
$[$20, 30$[$&  2697& 0.27&985273& 98.99\tabularnewline
$[$30, 40$[$&  2092& 0.21&987365& 99.20\tabularnewline
$[$40, 50$[$&  1680& 0.17&989045& 99.37\tabularnewline
$[$50, 100$[$&  4285& 0.43&993330& 99.80\tabularnewline
$[$100, 150$[$&  1263& 0.13&994593& 99.92\tabularnewline
$\geq$ 150&   767& 0.08&995360&100.00\tabularnewline
\hline
\end{tabular}\end{center}}

\end{table}

\begin{table}[h]
{\tiny
\caption{Distribution of the IP-weights for data-adaptive modeling approach for dynamic intervention $d_{7.0}$ and 15-day time-unit.\label{res}}
\begin{center}
\begin{tabular}{lllll}
\hline\hline
\multicolumn{1}{c}{Stabilized IPAW}&\multicolumn{1}{c}{Frequency}&\multicolumn{1}{c}{\%}&\multicolumn{1}{c}{Cumulative Frequency}&\multicolumn{1}{c}{Cumulative \%}\tabularnewline
\hline
\textless 0&      0& 0.00&      0&  0.00\tabularnewline
$[$0, 0.5$[$&1950648&98.51&1950648& 98.51\tabularnewline
$[$0.5, 1$[$&      0& 0.00&1950648& 98.51\tabularnewline
$[$1, 10$[$&   4917& 0.25&1955565& 98.76\tabularnewline
$[$10, 20$[$&   4060& 0.21&1959625& 98.96\tabularnewline
$[$20, 30$[$&   3435& 0.17&1963060& 99.14\tabularnewline
$[$30, 40$[$&   2748& 0.14&1965808& 99.28\tabularnewline
$[$40, 50$[$&   2331& 0.12&1968139& 99.39\tabularnewline
$[$50, 100$[$&   7186& 0.36&1975325& 99.76\tabularnewline
$[$100, 150$[$&   2650& 0.13&1977975& 99.89\tabularnewline
$\geq$ 150&   2182& 0.11&1980157&100.00\tabularnewline
\hline
\end{tabular}\end{center}}

\end{table}

\begin{table}[h]
{\tiny
\caption{Distribution of the IP-weights for data-adaptive modeling approach for dynamic intervention $d_{7.0}$ and 5-day time-unit.\label{res}}
\begin{center}
\begin{tabular}{lllll}
\hline\hline
\multicolumn{1}{c}{Stabilized IPAW}&\multicolumn{1}{c}{Frequency}&\multicolumn{1}{c}{\%}&\multicolumn{1}{c}{Cumulative Frequency}&\multicolumn{1}{c}{Cumulative \%}\tabularnewline
\hline
\textless 0&      0& 0.00&      0&  0.00\tabularnewline
$[$0, 0.5$[$&5885876&99.44&5885876& 99.44\tabularnewline
$[$0.5, 1$[$&      0& 0.00&5885876& 99.44\tabularnewline
$[$1, 10$[$&   1427& 0.02&5887303& 99.47\tabularnewline
$[$10, 20$[$&    182& 0.00&5887485& 99.47\tabularnewline
$[$20, 30$[$&     59& 0.00&5887544& 99.47\tabularnewline
$[$30, 40$[$&     80& 0.00&5887624& 99.47\tabularnewline
$[$40, 50$[$&    190& 0.00&5887814& 99.48\tabularnewline
$[$50, 100$[$&   5891& 0.10&5893705& 99.58\tabularnewline
$[$100, 150$[$&   6638& 0.11&5900343& 99.69\tabularnewline
$\geq$ 150&  18504& 0.31&5918847&100.00\tabularnewline
\hline
\end{tabular}\end{center}}

\end{table}

\begin{table}[h]
{\tiny
\caption{Distribution of the weights for data-adaptive modeling approach for dynamic intervention $d_{7.5}$ and 90-day time-unit.\label{res}}
\begin{center}
\begin{tabular}{lllll}
\hline\hline
\multicolumn{1}{c}{Stabilized IPAW}&\multicolumn{1}{c}{Frequency}&\multicolumn{1}{c}{\%}&\multicolumn{1}{c}{Cumulative Frequency}&\multicolumn{1}{c}{Cumulative \%}\tabularnewline
\hline
\textless 0&     0& 0.00&     0&  0.00\tabularnewline
$[$0, 0.5$[$&185363&54.71&185363& 54.71\tabularnewline
$[$0.5, 1$[$&     0& 0.00&185363& 54.71\tabularnewline
$[$1, 10$[$&148843&43.93&334206& 98.64\tabularnewline
$[$10, 20$[$&  3103& 0.92&337309& 99.56\tabularnewline
$[$20, 30$[$&   942& 0.28&338251& 99.84\tabularnewline
$[$30, 40$[$&   296& 0.09&338547& 99.92\tabularnewline
$[$40, 50$[$&   139& 0.04&338686& 99.96\tabularnewline
$[$50, 100$[$&   113& 0.03&338799&100.00\tabularnewline
$[$100, 150$[$&     7& 0.00&338806&100.00\tabularnewline
$\geq$ 150&     0& 0.00&338806&100.00\tabularnewline
\hline
\end{tabular}\end{center}}

\end{table}

\begin{table}[h]
{\tiny
\caption{Distribution of the weights for data-adaptive modeling approach for dynamic intervention $d_{7.5}$ and 30-day time-unit.\label{res}}
\begin{center}
\begin{tabular}{lllll}
\hline\hline
\multicolumn{1}{c}{Stabilized IPAW}&\multicolumn{1}{c}{Frequency}&\multicolumn{1}{c}{\%}&\multicolumn{1}{c}{Cumulative Frequency}&\multicolumn{1}{c}{Cumulative \%}\tabularnewline
\hline
\textless 0&     0& 0.00&     0&  0.00\tabularnewline
$[$0, 0.5$[$&561196&56.38&561196& 56.38\tabularnewline
$[$0.5, 1$[$&     0& 0.00&561196& 56.38\tabularnewline
$[$1, 10$[$&419483&42.14&980679& 98.53\tabularnewline
$[$10, 20$[$&  5846& 0.59&986525& 99.11\tabularnewline
$[$20, 30$[$&  3155& 0.32&989680& 99.43\tabularnewline
$[$30, 40$[$&  1912& 0.19&991592& 99.62\tabularnewline
$[$40, 50$[$&  1191& 0.12&992783& 99.74\tabularnewline
$[$50, 100$[$&  2005& 0.20&994788& 99.94\tabularnewline
$[$100, 150$[$&   369& 0.04&995157& 99.98\tabularnewline
$\geq$ 150&   203& 0.02&995360&100.00\tabularnewline
\hline
\end{tabular}\end{center}}

\end{table}

\begin{table}[h]
{\tiny
\caption{Distribution of the weights for data-adaptive modeling approach for dynamic intervention $d_{7.5}$ and 15-day time-unit.\label{res}}
\begin{center}
\begin{tabular}{lllll}
\hline\hline
\multicolumn{1}{c}{Stabilized IPAW}&\multicolumn{1}{c}{Frequency}&\multicolumn{1}{c}{\%}&\multicolumn{1}{c}{Cumulative Frequency}&\multicolumn{1}{c}{Cumulative \%}\tabularnewline
\hline
\textless 0&      0& 0.00&      0&  0.00\tabularnewline
$[$0, 0.5$[$&1128954&57.01&1128954& 57.01\tabularnewline
$[$0.5, 1$[$&      0& 0.00&1128954& 57.01\tabularnewline
$[$1, 10$[$& 828048&41.82&1957002& 98.83\tabularnewline
$[$10, 20$[$&   7370& 0.37&1964372& 99.20\tabularnewline
$[$20, 30$[$&   4462& 0.23&1968834& 99.43\tabularnewline
$[$30, 40$[$&   3119& 0.16&1971953& 99.59\tabularnewline
$[$40, 50$[$&   2267& 0.11&1974220& 99.70\tabularnewline
$[$50, 100$[$&   4502& 0.23&1978722& 99.93\tabularnewline
$[$100, 150$[$&    968& 0.05&1979690& 99.98\tabularnewline
$\geq$ 150&    467& 0.02&1980157&100.00\tabularnewline
\hline
\end{tabular}\end{center}}

\end{table}

\begin{table}[h]
{\tiny
\caption{Distribution of the weights for data-adaptive modeling approach for dynamic intervention $d_{7.5}$ and 5-day time-unit.\label{res}}
\begin{center}
\begin{tabular}{lllll}
\hline\hline
\multicolumn{1}{c}{Stabilized IPAW}&\multicolumn{1}{c}{Frequency}&\multicolumn{1}{c}{\%}&\multicolumn{1}{c}{Cumulative Frequency}&\multicolumn{1}{c}{Cumulative \%}\tabularnewline
\hline
\textless 0&      0& 0.00&      0&  0.00\tabularnewline
$[$0, 0.5$[$&3428305&57.92&3428305& 57.92\tabularnewline
$[$0.5, 1$[$&      0& 0.00&3428305& 57.92\tabularnewline
$[$1, 10$[$&2455423&41.48&5883728& 99.41\tabularnewline
$[$10, 20$[$&   5174& 0.09&5888902& 99.49\tabularnewline
$[$20, 30$[$&   2725& 0.05&5891627& 99.54\tabularnewline
$[$30, 40$[$&   1556& 0.03&5893183& 99.57\tabularnewline
$[$40, 50$[$&   1189& 0.02&5894372& 99.59\tabularnewline
$[$50, 100$[$&   9211& 0.16&5903583& 99.74\tabularnewline
$[$100, 150$[$&   6028& 0.10&5909611& 99.84\tabularnewline
$\geq$ 150&   9236& 0.16&5918847&100.00\tabularnewline
\hline
\end{tabular}\end{center}}

\end{table}

\begin{table}[h]
{\tiny
\caption{Distribution of the weights for data-adaptive modeling approach for dynamic intervention $d_{8.0}$ and 90-day time-unit.\label{res}}
\begin{center}
\begin{tabular}{lllll}
\hline\hline
\multicolumn{1}{c}{Stabilized IPAW}&\multicolumn{1}{c}{Frequency}&\multicolumn{1}{c}{\%}&\multicolumn{1}{c}{Cumulative Frequency}&\multicolumn{1}{c}{Cumulative \%}\tabularnewline
\hline
\textless 0&     0& 0.00&     0&  0.00\tabularnewline
$[$0, 0.5$[$&112225&33.12&112225& 33.12\tabularnewline
$[$0.5, 1$[$&     0& 0.00&112225& 33.12\tabularnewline
$[$1, 10$[$&224726&66.33&336951& 99.45\tabularnewline
$[$10, 20$[$&  1503& 0.44&338454& 99.90\tabularnewline
$[$20, 30$[$&   267& 0.08&338721& 99.97\tabularnewline
$[$30, 40$[$&    53& 0.02&338774& 99.99\tabularnewline
$[$40, 50$[$&    17& 0.01&338791&100.00\tabularnewline
$[$50, 100$[$&    15& 0.00&338806&100.00\tabularnewline
$[$100, 150$[$&     0& 0.00&338806&100.00\tabularnewline
$\geq$ 150&     0& 0.00&338806&100.00\tabularnewline
\hline
\end{tabular}\end{center}}

\end{table}

\begin{table}[h]
{\tiny
\caption{Distribution of the weights for data-adaptive modeling approach for dynamic intervention $d_{8.0}$ and 30-day time-unit.\label{res}}
\begin{center}
\begin{tabular}{lllll}
\hline\hline
\multicolumn{1}{c}{Stabilized IPAW}&\multicolumn{1}{c}{Frequency}&\multicolumn{1}{c}{\%}&\multicolumn{1}{c}{Cumulative Frequency}&\multicolumn{1}{c}{Cumulative \%}\tabularnewline
\hline
\textless 0&     0& 0.00&     0&  0.00\tabularnewline
$[$0, 0.5$[$&331832&33.34&331832& 33.34\tabularnewline
$[$0.5, 1$[$&     0& 0.00&331832& 33.34\tabularnewline
$[$1, 10$[$&654878&65.79&986710& 99.13\tabularnewline
$[$10, 20$[$&  4991& 0.50&991701& 99.63\tabularnewline
$[$20, 30$[$&  1861& 0.19&993562& 99.82\tabularnewline
$[$30, 40$[$&   853& 0.09&994415& 99.91\tabularnewline
$[$40, 50$[$&   424& 0.04&994839& 99.95\tabularnewline
$[$50, 100$[$&   477& 0.05&995316&100.00\tabularnewline
$[$100, 150$[$&    33& 0.00&995349&100.00\tabularnewline
$\geq$ 150&    11& 0.00&995360&100.00\tabularnewline
\hline
\end{tabular}\end{center}}

\end{table}

\begin{table}[h]
{\tiny
\caption{Distribution of the weights for data-adaptive modeling approach for dynamic intervention $d_{8.0}$ and 15-day time-unit.\label{res}}
\begin{center}
\begin{tabular}{lllll}
\hline\hline
\multicolumn{1}{c}{Stabilized IPAW}&\multicolumn{1}{c}{Frequency}&\multicolumn{1}{c}{\%}&\multicolumn{1}{c}{Cumulative Frequency}&\multicolumn{1}{c}{Cumulative \%}\tabularnewline
\hline
\textless 0&      0& 0.00&      0&  0.00\tabularnewline
$[$0, 0.5$[$& 664894&33.58& 664894& 33.58\tabularnewline
$[$0.5, 1$[$&      0& 0.00& 664894& 33.58\tabularnewline
$[$1, 10$[$&1299692&65.64&1964586& 99.21\tabularnewline
$[$10, 20$[$&   7211& 0.36&1971797& 99.58\tabularnewline
$[$20, 30$[$&   3336& 0.17&1975133& 99.75\tabularnewline
$[$30, 40$[$&   1967& 0.10&1977100& 99.85\tabularnewline
$[$40, 50$[$&   1099& 0.06&1978199& 99.90\tabularnewline
$[$50, 100$[$&   1682& 0.08&1979881& 99.99\tabularnewline
$[$100, 150$[$&    218& 0.01&1980099&100.00\tabularnewline
$\geq$ 150&     58& 0.00&1980157&100.00\tabularnewline
\hline
\end{tabular}\end{center}}

\end{table}

\begin{table}[h]
{\tiny
\caption{Distribution of the weights for data-adaptive modeling approach for dynamic intervention $d_{8.0}$ and 5-day time-unit.\label{res}}
\begin{center}
\begin{tabular}{lllll}
\hline\hline
\multicolumn{1}{c}{Stabilized IPAW}&\multicolumn{1}{c}{Frequency}&\multicolumn{1}{c}{\%}&\multicolumn{1}{c}{Cumulative Frequency}&\multicolumn{1}{c}{Cumulative \%}\tabularnewline
\hline
\textless 0&      0& 0.00&      0&  0.00\tabularnewline
$[$0, 0.5$[$&2016335&34.07&2016335& 34.07\tabularnewline
$[$0.5, 1$[$&      0& 0.00&2016335& 34.07\tabularnewline
$[$1, 10$[$&3874691&65.46&5891026& 99.53\tabularnewline
$[$10, 20$[$&   6688& 0.11&5897714& 99.64\tabularnewline
$[$20, 30$[$&   3224& 0.05&5900938& 99.70\tabularnewline
$[$30, 40$[$&   2009& 0.03&5902947& 99.73\tabularnewline
$[$40, 50$[$&   1410& 0.02&5904357& 99.76\tabularnewline
$[$50, 100$[$&   8103& 0.14&5912460& 99.89\tabularnewline
$[$100, 150$[$&   3512& 0.06&5915972& 99.95\tabularnewline
$\geq$ 150&   2875& 0.05&5918847&100.00\tabularnewline
\hline
\end{tabular}\end{center}}

\end{table}

\begin{table}[h]
{\tiny
\caption{Distribution of the weights for data-adaptive modeling approach for dynamic intervention $d_{8.5}$ and 90-day time-unit.\label{res}}
\begin{center}
\begin{tabular}{lllll}
\hline\hline
\multicolumn{1}{c}{Stabilized IPAW}&\multicolumn{1}{c}{Frequency}&\multicolumn{1}{c}{\%}&\multicolumn{1}{c}{Cumulative Frequency}&\multicolumn{1}{c}{Cumulative \%}\tabularnewline
\hline
\textless 0&     0& 0.00&     0&  0.00\tabularnewline
$[$0, 0.5$[$& 79414&23.44& 79414& 23.44\tabularnewline
$[$0.5, 1$[$&     0& 0.00& 79414& 23.44\tabularnewline
$[$1, 10$[$&258907&76.42&338321& 99.86\tabularnewline
$[$10, 20$[$&   436& 0.13&338757& 99.99\tabularnewline
$[$20, 30$[$&    38& 0.01&338795&100.00\tabularnewline
$[$30, 40$[$&     6& 0.00&338801&100.00\tabularnewline
$[$40, 50$[$&     4& 0.00&338805&100.00\tabularnewline
$[$50, 100$[$&     1& 0.00&338806&100.00\tabularnewline
$[$100, 150$[$&     0& 0.00&338806&100.00\tabularnewline
$\geq$ 150&     0& 0.00&338806&100.00\tabularnewline
\hline
\end{tabular}\end{center}}

\end{table}

\begin{table}[h]
{\tiny
\caption{Distribution of the weights for data-adaptive modeling approach for dynamic intervention $d_{8.5}$ and 30-day time-unit.\label{res}}
\begin{center}
\begin{tabular}{lllll}
\hline\hline
\multicolumn{1}{c}{Stabilized IPAW}&\multicolumn{1}{c}{Frequency}&\multicolumn{1}{c}{\%}&\multicolumn{1}{c}{Cumulative Frequency}&\multicolumn{1}{c}{Cumulative \%}\tabularnewline
\hline
\textless 0&     0& 0.00&     0&  0.00\tabularnewline
$[$0, 0.5$[$&220064&22.11&220064& 22.11\tabularnewline
$[$0.5, 1$[$&     0& 0.00&220064& 22.11\tabularnewline
$[$1, 10$[$&771661&77.53&991725& 99.63\tabularnewline
$[$10, 20$[$&  2631& 0.26&994356& 99.90\tabularnewline
$[$20, 30$[$&   626& 0.06&994982& 99.96\tabularnewline
$[$30, 40$[$&   219& 0.02&995201& 99.98\tabularnewline
$[$40, 50$[$&    78& 0.01&995279& 99.99\tabularnewline
$[$50, 100$[$&    64& 0.01&995343&100.00\tabularnewline
$[$100, 150$[$&     6& 0.00&995349&100.00\tabularnewline
$\geq$ 150&    11& 0.00&995360&100.00\tabularnewline
\hline
\end{tabular}\end{center}}

\end{table}

\begin{table}[h]
{\tiny
\caption{Distribution of the weights for data-adaptive modeling approach for dynamic intervention $d_{8.5}$ and 15-day time-unit.\label{res}}
\begin{center}
\begin{tabular}{lllll}
\hline\hline
\multicolumn{1}{c}{Stabilized IPAW}&\multicolumn{1}{c}{Frequency}&\multicolumn{1}{c}{\%}&\multicolumn{1}{c}{Cumulative Frequency}&\multicolumn{1}{c}{Cumulative \%}\tabularnewline
\hline
\textless 0&      0& 0.00&      0&  0.00\tabularnewline
$[$0, 0.5$[$& 433887&21.91& 433887& 21.91\tabularnewline
$[$0.5, 1$[$&      0& 0.00& 433887& 21.91\tabularnewline
$[$1, 10$[$&1538403&77.69&1972290& 99.60\tabularnewline
$[$10, 20$[$&   4972& 0.25&1977262& 99.85\tabularnewline
$[$20, 30$[$&   1421& 0.07&1978683& 99.93\tabularnewline
$[$30, 40$[$&    727& 0.04&1979410& 99.96\tabularnewline
$[$40, 50$[$&    355& 0.02&1979765& 99.98\tabularnewline
$[$50, 100$[$&    337& 0.02&1980102&100.00\tabularnewline
$[$100, 150$[$&     33& 0.00&1980135&100.00\tabularnewline
$\geq$ 150&     22& 0.00&1980157&100.00\tabularnewline
\hline
\end{tabular}\end{center}}

\end{table}

\begin{table}[h]
{\tiny
\caption{Distribution of the weights for data-adaptive modeling approach for dynamic intervention $d_{8.5}$ and 5-day time-unit.\label{res}}
\begin{center}
\begin{tabular}{lllll}
\hline\hline
\multicolumn{1}{c}{Stabilized IPAW}&\multicolumn{1}{c}{Frequency}&\multicolumn{1}{c}{\%}&\multicolumn{1}{c}{Cumulative Frequency}&\multicolumn{1}{c}{Cumulative \%}\tabularnewline
\hline
\textless 0&      0& 0.00&      0&  0.00\tabularnewline
$[$0, 0.5$[$&1300336&21.97&1300336& 21.97\tabularnewline
$[$0.5, 1$[$&      0& 0.00&1300336& 21.97\tabularnewline
$[$1, 10$[$&4601772&77.75&5902108& 99.72\tabularnewline
$[$10, 20$[$&   6470& 0.11&5908578& 99.83\tabularnewline
$[$20, 30$[$&   3342& 0.06&5911920& 99.88\tabularnewline
$[$30, 40$[$&   1717& 0.03&5913637& 99.91\tabularnewline
$[$40, 50$[$&   1048& 0.02&5914685& 99.93\tabularnewline
$[$50, 100$[$&   2812& 0.05&5917497& 99.98\tabularnewline
$[$100, 150$[$&    786& 0.01&5918283& 99.99\tabularnewline
$\geq$ 150&    564& 0.01&5918847&100.00\tabularnewline
\hline
\end{tabular}\end{center}}

\end{table}

\clearpage

\subsubsection*{Summary of the IP-weights for the parametric approach}

\begin{table}[h]
{\tiny
\caption{Distribution of the IP-weights for parametric modeling approach for dynamic intervention $d_{7.0}$ and 90-day time-unit.\label{res}}
\begin{center}
\begin{tabular}{lllll}
\hline\hline
\multicolumn{1}{c}{Stabilized IPAW}&\multicolumn{1}{c}{Frequency}&\multicolumn{1}{c}{\%}&\multicolumn{1}{c}{Cumulative Frequency}&\multicolumn{1}{c}{Cumulative \%}\tabularnewline
\hline
\textless 0&     0& 0.00&     0&  0.00\tabularnewline
$[$0, 0.5$[$&326730&96.44&326730& 96.44\tabularnewline
$[$0.5, 1$[$&     0& 0.00&326730& 96.44\tabularnewline
$[$1, 10$[$&  1916& 0.57&328646& 97.00\tabularnewline
$[$10, 20$[$&  3362& 0.99&332008& 97.99\tabularnewline
$[$20, 30$[$&  2277& 0.67&334285& 98.67\tabularnewline
$[$30, 40$[$&  1458& 0.43&335743& 99.10\tabularnewline
$[$40, 50$[$&   952& 0.28&336695& 99.38\tabularnewline
$[$50, 100$[$&  1655& 0.49&338350& 99.87\tabularnewline
$[$100, 150$[$&   310& 0.09&338660& 99.96\tabularnewline
$\geq$ 150&   146& 0.04&338806&100.00\tabularnewline
\hline
\end{tabular}\end{center}}

\end{table}

\begin{table}[h]
{\tiny
\caption{Distribution of the IP-weights for parametric modeling approach for dynamic intervention $d_{7.0}$ and 30-day time-unit.\label{res}}
\begin{center}
\begin{tabular}{lllll}
\hline\hline
\multicolumn{1}{c}{Stabilized IPAW}&\multicolumn{1}{c}{Frequency}&\multicolumn{1}{c}{\%}&\multicolumn{1}{c}{Cumulative Frequency}&\multicolumn{1}{c}{Cumulative \%}\tabularnewline
\hline
\textless 0&     0& 0.00&     0&  0.00\tabularnewline
$[$0, 0.5$[$&974355&97.89&974355& 97.89\tabularnewline
$[$0.5, 1$[$&     0& 0.00&974355& 97.89\tabularnewline
$[$1, 10$[$&   830& 0.08&975185& 97.97\tabularnewline
$[$10, 20$[$&  2969& 0.30&978154& 98.27\tabularnewline
$[$20, 30$[$&  3316& 0.33&981470& 98.60\tabularnewline
$[$30, 40$[$&  2854& 0.29&984324& 98.89\tabularnewline
$[$40, 50$[$&  2243& 0.23&986567& 99.12\tabularnewline
$[$50, 100$[$&  5537& 0.56&992104& 99.67\tabularnewline
$[$100, 150$[$&  1795& 0.18&993899& 99.85\tabularnewline
$\geq$ 150&  1461& 0.15&995360&100.00\tabularnewline
\hline
\end{tabular}\end{center}}

\end{table}

\begin{table}[h]
{\tiny
\caption{Distribution of the IP-weights for parametric modeling approach for dynamic intervention $d_{7.0}$ and 15-day time-unit.\label{res}}
\begin{center}
\begin{tabular}{lllll}
\hline\hline
\multicolumn{1}{c}{Stabilized IPAW}&\multicolumn{1}{c}{Frequency}&\multicolumn{1}{c}{\%}&\multicolumn{1}{c}{Cumulative Frequency}&\multicolumn{1}{c}{Cumulative \%}\tabularnewline
\hline
\textless 0&      0& 0.00&      0&  0.00\tabularnewline
$[$0, 0.5$[$&1950648&98.51&1950648& 98.51\tabularnewline
$[$0.5, 1$[$&      0& 0.00&1950648& 98.51\tabularnewline
$[$1, 10$[$&    581& 0.03&1951229& 98.54\tabularnewline
$[$10, 20$[$&   2020& 0.10&1953249& 98.64\tabularnewline
$[$20, 30$[$&   3148& 0.16&1956397& 98.80\tabularnewline
$[$30, 40$[$&   3127& 0.16&1959524& 98.96\tabularnewline
$[$40, 50$[$&   2885& 0.15&1962409& 99.10\tabularnewline
$[$50, 100$[$&   9412& 0.48&1971821& 99.58\tabularnewline
$[$100, 150$[$&   3918& 0.20&1975739& 99.78\tabularnewline
$\geq$ 150&   4418& 0.22&1980157&100.00\tabularnewline
\hline
\end{tabular}\end{center}}

\end{table}

\begin{table}[h]
{\tiny
\caption{Distribution of the IP-weights for parametric modeling approach for dynamic intervention $d_{7.0}$ and 5-day time-unit.\label{res}}
\begin{center}
\begin{tabular}{lllll}
\hline\hline
\multicolumn{1}{c}{Stabilized IPAW}&\multicolumn{1}{c}{Frequency}&\multicolumn{1}{c}{\%}&\multicolumn{1}{c}{Cumulative Frequency}&\multicolumn{1}{c}{Cumulative \%}\tabularnewline
\hline
\textless 0&      0& 0.00&      0&  0.00\tabularnewline
$[$0, 0.5$[$&5885876&99.44&5885876& 99.44\tabularnewline
$[$0.5, 1$[$&      0& 0.00&5885876& 99.44\tabularnewline
$[$1, 10$[$&    359& 0.01&5886235& 99.45\tabularnewline
$[$10, 20$[$&    382& 0.01&5886617& 99.46\tabularnewline
$[$20, 30$[$&    683& 0.01&5887300& 99.47\tabularnewline
$[$30, 40$[$&    964& 0.02&5888264& 99.48\tabularnewline
$[$40, 50$[$&   1321& 0.02&5889585& 99.51\tabularnewline
$[$50, 100$[$&   7019& 0.12&5896604& 99.62\tabularnewline
$[$100, 150$[$&   5762& 0.10&5902366& 99.72\tabularnewline
$\geq$ 150&  16481& 0.28&5918847&100.00\tabularnewline
\hline
\end{tabular}\end{center}}

\end{table}

\begin{table}[h]
{\tiny
\caption{Distribution of the weights for parametric modeling approach for dynamic intervention $d_{7.5}$ and 90-day time-unit.\label{res}}
\begin{center}
\begin{tabular}{lllll}
\hline\hline
\multicolumn{1}{c}{Stabilized IPAW}&\multicolumn{1}{c}{Frequency}&\multicolumn{1}{c}{\%}&\multicolumn{1}{c}{Cumulative Frequency}&\multicolumn{1}{c}{Cumulative \%}\tabularnewline
\hline
\textless 0&     0& 0.00&     0&  0.00\tabularnewline
$[$0, 0.5$[$&185363&54.71&185363& 54.71\tabularnewline
$[$0.5, 1$[$&     0& 0.00&185363& 54.71\tabularnewline
$[$1, 10$[$&145121&42.83&330484& 97.54\tabularnewline
$[$10, 20$[$&  4438& 1.31&334922& 98.85\tabularnewline
$[$20, 30$[$&  1957& 0.58&336879& 99.43\tabularnewline
$[$30, 40$[$&   855& 0.25&337734& 99.68\tabularnewline
$[$40, 50$[$&   420& 0.12&338154& 99.81\tabularnewline
$[$50, 100$[$&   569& 0.17&338723& 99.98\tabularnewline
$[$100, 150$[$&    62& 0.02&338785& 99.99\tabularnewline
$\geq$ 150&    21& 0.01&338806&100.00\tabularnewline
\hline
\end{tabular}\end{center}}

\end{table}

\begin{table}[h]
{\tiny
\caption{Distribution of the weights for parametric modeling approach for dynamic intervention $d_{7.5}$ and 30-day time-unit.\label{res}}
\begin{center}
\begin{tabular}{lllll}
\hline\hline
\multicolumn{1}{c}{Stabilized IPAW}&\multicolumn{1}{c}{Frequency}&\multicolumn{1}{c}{\%}&\multicolumn{1}{c}{Cumulative Frequency}&\multicolumn{1}{c}{Cumulative \%}\tabularnewline
\hline
\textless 0&     0& 0.00&     0&  0.00\tabularnewline
$[$0, 0.5$[$&561196&56.38&561196& 56.38\tabularnewline
$[$0.5, 1$[$&     0& 0.00&561196& 56.38\tabularnewline
$[$1, 10$[$&414926&41.69&976122& 98.07\tabularnewline
$[$10, 20$[$&  5369& 0.54&981491& 98.61\tabularnewline
$[$20, 30$[$&  4088& 0.41&985579& 99.02\tabularnewline
$[$30, 40$[$&  2845& 0.29&988424& 99.30\tabularnewline
$[$40, 50$[$&  1964& 0.20&990388& 99.50\tabularnewline
$[$50, 100$[$&  3477& 0.35&993865& 99.85\tabularnewline
$[$100, 150$[$&   889& 0.09&994754& 99.94\tabularnewline
$\geq$ 150&   606& 0.06&995360&100.00\tabularnewline
\hline
\end{tabular}\end{center}}

\end{table}

\begin{table}[h]
{\tiny
\caption{Distribution of the weights for parametric modeling approach for dynamic intervention $d_{7.5}$ and 15-day time-unit.\label{res}}
\begin{center}
\begin{tabular}{lllll}
\hline\hline
\multicolumn{1}{c}{Stabilized IPAW}&\multicolumn{1}{c}{Frequency}&\multicolumn{1}{c}{\%}&\multicolumn{1}{c}{Cumulative Frequency}&\multicolumn{1}{c}{Cumulative \%}\tabularnewline
\hline
\textless 0&      0& 0.00&      0&  0.00\tabularnewline
$[$0, 0.5$[$&1128954&57.01&1128954& 57.01\tabularnewline
$[$0.5, 1$[$&      0& 0.00&1128954& 57.01\tabularnewline
$[$1, 10$[$& 822710&41.55&1951664& 98.56\tabularnewline
$[$10, 20$[$&   5460& 0.28&1957124& 98.84\tabularnewline
$[$20, 30$[$&   4502& 0.23&1961626& 99.06\tabularnewline
$[$30, 40$[$&   3582& 0.18&1965208& 99.25\tabularnewline
$[$40, 50$[$&   2839& 0.14&1968047& 99.39\tabularnewline
$[$50, 100$[$&   7565& 0.38&1975612& 99.77\tabularnewline
$[$100, 150$[$&   2376& 0.12&1977988& 99.89\tabularnewline
$\geq$ 150&   2169& 0.11&1980157&100.00\tabularnewline
\hline
\end{tabular}\end{center}}

\end{table}

\begin{table}[h]
{\tiny
\caption{Distribution of the weights for parametric modeling approach for dynamic intervention $d_{7.5}$ and 5-day time-unit.\label{res}}
\begin{center}
\begin{tabular}{lllll}
\hline\hline
\multicolumn{1}{c}{Stabilized IPAW}&\multicolumn{1}{c}{Frequency}&\multicolumn{1}{c}{\%}&\multicolumn{1}{c}{Cumulative Frequency}&\multicolumn{1}{c}{Cumulative \%}\tabularnewline
\hline
\textless 0&      0& 0.00&      0&  0.00\tabularnewline
$[$0, 0.5$[$&3428305&57.92&3428305& 57.92\tabularnewline
$[$0.5, 1$[$&      0& 0.00&3428305& 57.92\tabularnewline
$[$1, 10$[$&2453172&41.45&5881477& 99.37\tabularnewline
$[$10, 20$[$&   4277& 0.07&5885754& 99.44\tabularnewline
$[$20, 30$[$&   2882& 0.05&5888636& 99.49\tabularnewline
$[$30, 40$[$&   2021& 0.03&5890657& 99.52\tabularnewline
$[$40, 50$[$&   1966& 0.03&5892623& 99.56\tabularnewline
$[$50, 100$[$&   7641& 0.13&5900264& 99.69\tabularnewline
$[$100, 150$[$&   5446& 0.09&5905710& 99.78\tabularnewline
$\geq$ 150&  13137& 0.22&5918847&100.00\tabularnewline
\hline
\end{tabular}\end{center}}

\end{table}

\begin{table}[h]
{\tiny
\caption{Distribution of the weights for parametric modeling approach for dynamic intervention $d_{8.0}$ and 90-day time-unit.\label{res}}
\begin{center}
\begin{tabular}{lllll}
\hline\hline
\multicolumn{1}{c}{Stabilized IPAW}&\multicolumn{1}{c}{Frequency}&\multicolumn{1}{c}{\%}&\multicolumn{1}{c}{Cumulative Frequency}&\multicolumn{1}{c}{Cumulative \%}\tabularnewline
\hline
\textless 0&     0& 0.00&     0&  0.00\tabularnewline
$[$0, 0.5$[$&112225&33.12&112225& 33.12\tabularnewline
$[$0.5, 1$[$&     0& 0.00&112225& 33.12\tabularnewline
$[$1, 10$[$&222181&65.58&334406& 98.70\tabularnewline
$[$10, 20$[$&  3083& 0.91&337489& 99.61\tabularnewline
$[$20, 30$[$&   861& 0.25&338350& 99.87\tabularnewline
$[$30, 40$[$&   252& 0.07&338602& 99.94\tabularnewline
$[$40, 50$[$&    95& 0.03&338697& 99.97\tabularnewline
$[$50, 100$[$&   102& 0.03&338799&100.00\tabularnewline
$[$100, 150$[$&     7& 0.00&338806&100.00\tabularnewline
$\geq$ 150&     0& 0.00&338806&100.00\tabularnewline
\hline
\end{tabular}\end{center}}

\end{table}

\begin{table}[h]
{\tiny
\caption{Distribution of the weights for parametric modeling approach for dynamic intervention $d_{8.0}$ and 30-day time-unit.\label{res}}
\begin{center}
\begin{tabular}{lllll}
\hline\hline
\multicolumn{1}{c}{Stabilized IPAW}&\multicolumn{1}{c}{Frequency}&\multicolumn{1}{c}{\%}&\multicolumn{1}{c}{Cumulative Frequency}&\multicolumn{1}{c}{Cumulative \%}\tabularnewline
\hline
\textless 0&     0& 0.00&     0&  0.00\tabularnewline
$[$0, 0.5$[$&331832&33.34&331832& 33.34\tabularnewline
$[$0.5, 1$[$&     0& 0.00&331832& 33.34\tabularnewline
$[$1, 10$[$&650472&65.35&982304& 98.69\tabularnewline
$[$10, 20$[$&  5463& 0.55&987767& 99.24\tabularnewline
$[$20, 30$[$&  3077& 0.31&990844& 99.55\tabularnewline
$[$30, 40$[$&  1779& 0.18&992623& 99.73\tabularnewline
$[$40, 50$[$&  1053& 0.11&993676& 99.83\tabularnewline
$[$50, 100$[$&  1344& 0.14&995020& 99.97\tabularnewline
$[$100, 150$[$&   236& 0.02&995256& 99.99\tabularnewline
$\geq$ 150&   104& 0.01&995360&100.00\tabularnewline
\hline
\end{tabular}\end{center}}

\end{table}

\begin{table}[h]
{\tiny
\caption{Distribution of the weights for parametric modeling approach for dynamic intervention $d_{8.0}$ and 15-day time-unit.\label{res}}
\begin{center}
\begin{tabular}{lllll}
\hline\hline
\multicolumn{1}{c}{Stabilized IPAW}&\multicolumn{1}{c}{Frequency}&\multicolumn{1}{c}{\%}&\multicolumn{1}{c}{Cumulative Frequency}&\multicolumn{1}{c}{Cumulative \%}\tabularnewline
\hline
\textless 0&      0& 0.00&      0&  0.00\tabularnewline
$[$0, 0.5$[$& 664894&33.58& 664894& 33.58\tabularnewline
$[$0.5, 1$[$&      0& 0.00& 664894& 33.58\tabularnewline
$[$1, 10$[$&1294941&65.40&1959835& 98.97\tabularnewline
$[$10, 20$[$&   5850& 0.30&1965685& 99.27\tabularnewline
$[$20, 30$[$&   3828& 0.19&1969513& 99.46\tabularnewline
$[$30, 40$[$&   2744& 0.14&1972257& 99.60\tabularnewline
$[$40, 50$[$&   2001& 0.10&1974258& 99.70\tabularnewline
$[$50, 100$[$&   4340& 0.22&1978598& 99.92\tabularnewline
$[$100, 150$[$&    974& 0.05&1979572& 99.97\tabularnewline
$\geq$ 150&    585& 0.03&1980157&100.00\tabularnewline
\hline
\end{tabular}\end{center}}

\end{table}

\begin{table}[h]
{\tiny
\caption{Distribution of the weights for parametric modeling approach for dynamic intervention $d_{8.0}$ and 5-day time-unit.\label{res}}
\begin{center}
\begin{tabular}{lllll}
\hline\hline
\multicolumn{1}{c}{Stabilized IPAW}&\multicolumn{1}{c}{Frequency}&\multicolumn{1}{c}{\%}&\multicolumn{1}{c}{Cumulative Frequency}&\multicolumn{1}{c}{Cumulative \%}\tabularnewline
\hline
\textless 0&      0& 0.00&      0&  0.00\tabularnewline
$[$0, 0.5$[$&2016335&34.07&2016335& 34.07\tabularnewline
$[$0.5, 1$[$&      0& 0.00&2016335& 34.07\tabularnewline
$[$1, 10$[$&3872051&65.42&5888386& 99.49\tabularnewline
$[$10, 20$[$&   5028& 0.08&5893414& 99.57\tabularnewline
$[$20, 30$[$&   2992& 0.05&5896406& 99.62\tabularnewline
$[$30, 40$[$&   2033& 0.03&5898439& 99.66\tabularnewline
$[$40, 50$[$&   1668& 0.03&5900107& 99.68\tabularnewline
$[$50, 100$[$&   6565& 0.11&5906672& 99.79\tabularnewline
$[$100, 150$[$&   4233& 0.07&5910905& 99.87\tabularnewline
$\geq$ 150&   7942& 0.13&5918847&100.00\tabularnewline
\hline
\end{tabular}\end{center}}

\end{table}

\begin{table}[h]
{\tiny
\caption{Distribution of the weights for parametric modeling approach for dynamic intervention $d_{8.5}$ and 90-day time-unit.\label{res}}
\begin{center}
\begin{tabular}{lllll}
\hline\hline
\multicolumn{1}{c}{Stabilized IPAW}&\multicolumn{1}{c}{Frequency}&\multicolumn{1}{c}{\%}&\multicolumn{1}{c}{Cumulative Frequency}&\multicolumn{1}{c}{Cumulative \%}\tabularnewline
\hline
\textless 0&     0& 0.00&     0&  0.00\tabularnewline
$[$0, 0.5$[$& 79414&23.44& 79414& 23.44\tabularnewline
$[$0.5, 1$[$&     0& 0.00& 79414& 23.44\tabularnewline
$[$1, 10$[$&257975&76.14&337389& 99.58\tabularnewline
$[$10, 20$[$&  1153& 0.34&338542& 99.92\tabularnewline
$[$20, 30$[$&   194& 0.06&338736& 99.98\tabularnewline
$[$30, 40$[$&    40& 0.01&338776& 99.99\tabularnewline
$[$40, 50$[$&    15& 0.00&338791&100.00\tabularnewline
$[$50, 100$[$&    13& 0.00&338804&100.00\tabularnewline
$[$100, 150$[$&     2& 0.00&338806&100.00\tabularnewline
$\geq$ 150&     0& 0.00&338806&100.00\tabularnewline
\hline
\end{tabular}\end{center}}

\end{table}

\begin{table}[h]
{\tiny
\caption{Distribution of the weights for parametric modeling approach for dynamic intervention $d_{8.5}$ and 30-day time-unit.\label{res}}
\begin{center}
\begin{tabular}{lllll}
\hline\hline
\multicolumn{1}{c}{Stabilized IPAW}&\multicolumn{1}{c}{Frequency}&\multicolumn{1}{c}{\%}&\multicolumn{1}{c}{Cumulative Frequency}&\multicolumn{1}{c}{Cumulative \%}\tabularnewline
\hline
\textless 0&     0& 0.00&     0&  0.00\tabularnewline
$[$0, 0.5$[$&220064&22.11&220064& 22.11\tabularnewline
$[$0.5, 1$[$&     0& 0.00&220064& 22.11\tabularnewline
$[$1, 10$[$&769260&77.28&989324& 99.39\tabularnewline
$[$10, 20$[$&  3421& 0.34&992745& 99.74\tabularnewline
$[$20, 30$[$&  1385& 0.14&994130& 99.88\tabularnewline
$[$30, 40$[$&   586& 0.06&994716& 99.94\tabularnewline
$[$40, 50$[$&   288& 0.03&995004& 99.96\tabularnewline
$[$50, 100$[$&   285& 0.03&995289& 99.99\tabularnewline
$[$100, 150$[$&    45& 0.00&995334&100.00\tabularnewline
$\geq$ 150&    26& 0.00&995360&100.00\tabularnewline
\hline
\end{tabular}\end{center}}

\end{table}

\begin{table}[h]
{\tiny
\caption{Distribution of the weights for parametric modeling approach for dynamic intervention $d_{8.5}$ and 15-day time-unit.\label{res}}
\begin{center}
\begin{tabular}{lllll}
\hline\hline
\multicolumn{1}{c}{Stabilized IPAW}&\multicolumn{1}{c}{Frequency}&\multicolumn{1}{c}{\%}&\multicolumn{1}{c}{Cumulative Frequency}&\multicolumn{1}{c}{Cumulative \%}\tabularnewline
\hline
\textless 0&      0& 0.00&      0&  0.00\tabularnewline
$[$0, 0.5$[$& 433887&21.91& 433887& 21.91\tabularnewline
$[$0.5, 1$[$&      0& 0.00& 433887& 21.91\tabularnewline
$[$1, 10$[$&1535888&77.56&1969775& 99.48\tabularnewline
$[$10, 20$[$&   4079& 0.21&1973854& 99.68\tabularnewline
$[$20, 30$[$&   2242& 0.11&1976096& 99.79\tabularnewline
$[$30, 40$[$&   1554& 0.08&1977650& 99.87\tabularnewline
$[$40, 50$[$&    882& 0.04&1978532& 99.92\tabularnewline
$[$50, 100$[$&   1252& 0.06&1979784& 99.98\tabularnewline
$[$100, 150$[$&    223& 0.01&1980007& 99.99\tabularnewline
$\geq$ 150&    150& 0.01&1980157&100.00\tabularnewline
\hline
\end{tabular}\end{center}}

\end{table}

\begin{table}[h]
{\tiny
\caption{Distribution of the weights for parametric modeling approach for dynamic intervention $d_{8.5}$ and 5-day time-unit.\label{res}}
\begin{center}
\begin{tabular}{lllll}
\hline\hline
\multicolumn{1}{c}{Stabilized IPAW}&\multicolumn{1}{c}{Frequency}&\multicolumn{1}{c}{\%}&\multicolumn{1}{c}{Cumulative Frequency}&\multicolumn{1}{c}{Cumulative \%}\tabularnewline
\hline
\textless 0&      0& 0.00&      0&  0.00\tabularnewline
$[$0, 0.5$[$&1300336&21.97&1300336& 21.97\tabularnewline
$[$0.5, 1$[$&      0& 0.00&1300336& 21.97\tabularnewline
$[$1, 10$[$&4601590&77.74&5901926& 99.71\tabularnewline
$[$10, 20$[$&   3994& 0.07&5905920& 99.78\tabularnewline
$[$20, 30$[$&   1973& 0.03&5907893& 99.81\tabularnewline
$[$30, 40$[$&   1396& 0.02&5909289& 99.84\tabularnewline
$[$40, 50$[$&   1192& 0.02&5910481& 99.86\tabularnewline
$[$50, 100$[$&   4206& 0.07&5914687& 99.93\tabularnewline
$[$100, 150$[$&   1895& 0.03&5916582& 99.96\tabularnewline
$\geq$ 150&   2265& 0.04&5918847&100.00\tabularnewline
\hline
\end{tabular}\end{center}}

\end{table}

\clearpage

\bibliographystyle{biom}
\bibliography{LTMLE_stremr_diabetes_study}

\end{document}